\documentclass[
twocolumn,
 prb,
 amsmath,
 amssymb,
]{revtex4}

\usepackage{graphicx}
\usepackage{dcolumn}
\usepackage{bm}
\usepackage{relsize}
\usepackage{subfig}

\preprint{AIP/123-QED}

\begin{document}

\title{A theoretical model for parallel SQUID arrays with fluxoid focussing} 

\author{K.-H. M\"{u}ller}
\email{karl.muller@csiro.au}

\author{E. E. Mitchell}%

\affiliation{CSIRO Manufacturing, PO Box 218, Lindfield, NSW 2070, Australia
}%

\date{\today}

\begin{abstract}
We have developed a comprehensive theoretical model for predicting the magnetic field response of a parallel SQUID array in the voltage state. The model predictions are compared with our experimental data from a parallel SQUID array made of a YBCO thin-film patterned into wide tracks, busbars and leads, with eleven  step-edge Josephson junctions. Our theoretical model uses the Josephson equations for resistively shunted junctions as well as the second Ginzburg-Landau equation to derive a system of coupled first-order nonlinear differential equations to describe the time-evolution of the Josephson junction phase differences which includes Johnson noise. Employing the second London equation and Biot-Savart's law, the supercurrent density distribution is calculated, using the stream function approach, which leads to a 2D second-order linear Fredholm integro-differential equation for the stream function with time-dependent boundary conditions. The novelty of the model is that it calculates the stream function everywhere in the thin-film structure to determine during the time-evolution the fluxoids for each SQUID array hole. Our numerical model calculations are compared with our experimental data and predict the bias-current versus voltage and the voltage versus magnetic field response with unprecedented accuracy. The model elucidates the importance to fully take Meissner shielding and current crowding into account in order to properly describe fluxoid focussing and bias-current injection. Furthermore, our model illustrates the failure of the simple lumped-element approach to describe a parallel SQUID array with a wide thin-film structure.
%
\end{abstract}

\maketitle

%
\section{\label{sec:level1}Introduction}
In this paper we develop a comprehensive theoretical model for a parallel SQUID array with over-damped Josephson junctions (JJs), made from a high-$T_c$ superconducting thin-film with wide tracks, busbars and leads, exposed to an applied magnetic field while driven by a bias-current into the voltage state. Opposite to the commonly used lumped-element model, our model can describe for the first time the experimental data of a parallel SQUID array with unprecedented accuracy. Highly accurate predictions are important for the optimisation of magnetometers, low noise current amplifiers and high frequency AC magnetic field sensors.\\
\par
Enhanced quantum interference in a parallel SQUID array was first mentioned by Feynman {\it{et al.}} \cite{FEY66}, and was predicted to show up as a sharpening of the array's critical current peak, seen in measurements of the array's critical current as a function of an applied magnetic field. Such an enhancement was observed experimentally for the first time in the 1960s in parallel arrays of superconducting point contacts \cite{ZIM66, WAE68}. A three-junction parallel SQUID array was fabricated and theoretically described in the early 1970s using a simple lumped-element model \cite{STU72}. Later a parallel SQUID array (named superconducting quantum interference grating (SQUIG)) with 11 JJs was theoretically modelled with an improved lumped-element model by Miller {\it{et al.}} \cite{MIL91}, revealing in detail the mechanism of the enhanced quantum interference effect and its dependence on the SQUID array's screening parameter $\beta_L$.
In the late 1980s and early 1990s interest in modelling of two-dimensional (2D) arrays of JJs emerged, due to the realisation that most HTS materials are granular where grain boundaries act as JJs \cite{MUL89}. In contrast to parallel SQUID arrays which only have vertical JJ connections, 2D JJ-arrays have both vertical and horizontal JJ connections which form the plaquettes of the array \cite{LOB00}. In the early 2000s parallel SQUID arrays with varying hole area sizes between junctions, named superconducting quantum interference filters (SQIFs), were investigated experimentally and theoretically, again using a lumped-element model \cite{OPP01, OPP02}.
Serial SQUID arrays were also studied \cite{HAU01, LON11a} and the question of how to optimise linearity was addressed \cite{KOR09, KOR09a, LON11}. Furthermore, the performance of parallel SQIF arrays put in series (2D SQIF arrays) have also been investigated by Kornev {\it{et al.}} \cite{KOR11}, Taylor {\it{et al.}}\, \cite{TAY16} and Mitchell {\it{et al.}}  \, \cite{MIT16}. In a recent review, theoretical and experimental studies of different SQUID arrays made from LTS materials and used as miniature antennas have been compared \cite{KOR17}.
The similarities between interference patterns in parallel SQUID arrays and optical multiple slit gratings has been discussed by De Luca \cite{DEL15}. Last year one and two dimensional SQUID arrays have been investigated further experimentally and theoretically \cite{CHO19, CHE19, MIT19}. In addition, a review about design tools and progress in modelling of superconducting circuits was written by Fourie \cite{FOR18}. \\
\par
The often used lumped-element model can only be applied if a SQUID array consists of sufficiently narrow superonducting tracks such that Kirchhoff's law can be applied at well defined current vertices \cite{OPP01, OPP02}. However, SQUID arrays are usually made from thin-film structures with wide tracks, busbars and leads, where the Meissner shielding from wide superconducting structures creates strong magnetic flux-focussing and current crowding. Neither flux-focussing nor current crowding are not part of the lumped-element model. Going beyond the lumped-element model, a two junction SQUID array (the normal DC SQUID) with a wide washer structure in the zero voltage state has been investigated theoretically in an approximate way by Clem and  Brandt \cite{CLE05}. Terauchi {\it{et al.}} \cite{TER15} investigate the effect of wider tracks on the shape of the voltage pulses of a DC SQUID. \\
\par
Here, in our paper, we have developed a comprehensive theoretical model for parallel SQUID arrays in the voltage state. In particular, we consider wide superconducting  thin-film structures in the Meissner state and incorporate into the Josephson array equations the time-dependent supercurrent density distribution of the array, obtained from the second London equation and Biot-Savart's law. In contrast to the lumped-element model, our model does not make use of any lumped-element inductances but instead calculates the values for the fluxoids of each hole in the array during the time-evolution of the JJ phase differences. The static supercurrent density and magnetic field distributions in different superconducting thin-film geometrical structures, based on London and Maxwell equations, with and without DC current injection,  but without any JJ's, has been studied previously \cite{BRA05, BRA07, KHA01, KHA03}.\\
\par
Our paper is organised as follows. In Sec. II we outline in detail our theoretical model for parallel SQUID arrays with wide superconducting thin-film structures. In Sec. III we briefly mention our device fabrication and experimental set up. In Sec. IV we present the results of our model calculations and compare them with our YBCO thin-film array experimental data as well as with results from a lumped-element model calculation. We summarise our findings in Sec. V.\\

\section{\label{sec:level1}Theorectical model  for parallel SQUID arrays}

The main goal of this paper is to calculate the voltage response of a thin-film parallel SQUID array to an applied perpendicular magnetic field and compare our results with our experimental data for a YBCO thin-film parallel SQUID array. Contrary to the commonly used lumped-element model \cite{MIL91, OPP01}, we will use the second London equation and Biot-Savart's law to calculate from the supercurrent density within the array the fluxoids of the SQUID array holes during the oscillatory time-evolution of the array \cite{CLE05}.\\

\begin{figure}[h]
\begin{center}
\includegraphics[width=0.5\textwidth]{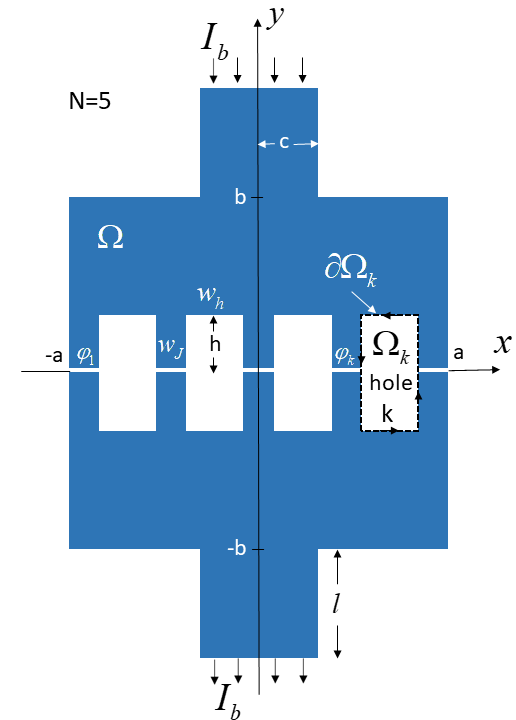}
\caption{Example of a parallel thin-film parallel SQUID array with $N = 5$ JJs. The domain $\Omega$ (blue) is made of a superconducting YBCO thin-film. All JJs have the same width $w_J$, and the hole domains $\Omega_k$ are of the same width $w_h$ and length $2 h$. $\partial \Omega_k$ is the boundary of hole $k$. $\varphi_k$ are the gauge invariant phase differences across the JJs, where $k=1, ... , N$. The JJs are connected via wide superconducting horizontal busbars on the top and bottom, each of area $2a (b-h)$ with wide attached superconducting leads, $2 c$ wide and $l$ long, and $w_J$ wide tracks attached to the JJs. In our fabricated devices, $l$ is much longer than shown here. A bias-current $I_b$ is injected into the top lead and exits from the bottom lead. The time-averaged voltage $V$ is measured between the ends of the two leads.}
\label{default}
\end{center}
\end{figure}
\par

As an example, Fig.1 displays a planar thin-film parallel SQUID array with $N$ = 5 JJs and $N-1$ holes, with wide tracks, busbars and bias-current leads. The array lies in the $xy$ plane and the magnetic induction $B_a$ is applied perpendicular in $z$ direction. The array is symmetric about both the $x$ and $y$ axis. \\

\subsection{\label{sec:level2}Josephson junction phase differences and fluxoids}

In our case the width $w_J$ of the JJs is much less than the Josephson penetration depth $\lambda_J$ \cite{TOL96} (short junction) and the applied magnetic induction $B_a$ is sufficiently small such that the JJ current density is nearly constant across junction areas. In this case, the current across the junctions is described by the Josephson equation $I_{ck} \sin \varphi_k(t)$ \cite{JOS69}, where $\varphi_k(t)$ is the gauge-invariant phase difference across the $k^{\text{th}}$ junction at time $t$ and $I_{ck}$ the junction critical current with $k$=1,...,$N$.\\
\par
Because in high-temperature superconducting materials, like YBCO thin-films, the capacitance of the fabricated JJs is small, one can apply the resistively shunted junction (RSJ) model to describe the time-dependent total current $I_k(t)$ 
flowing through the junctions,
 {\it{i.e.}}
\begin{equation}
I_k(t) =  \frac{V_k(t)}{R_k} + I_{ck}  \sin \varphi_k(t) + I_k^{Noise}(t)  \; .
\end{equation}
Here $R_k$ is the normal resistance of the $k^{\text{th}}$ junction and $V_k(t)$ the voltage across that junction, which according to the Josephson equation is 
\begin{equation}
V_k(t)=\frac{\Phi_0}{2\pi} \, \frac{d \varphi_k(t)}{d t} \; ,
\end{equation}   
where $\Phi_0$ is the flux quantum.
$I_k^{Noise}(t)$ in Eq.(1) is the Johnson noise current, originating at finite temperature  from the resistor $R_k$. This noise is often also called Nyquist noise or white noise.\\
\par
From the second Ginzburg-Landau equation \cite{TIN04} it follows that 
\begin{equation}
\varphi_{k+1} (t)- \varphi_k(t) = \frac{2 \pi} {\Phi_0} \, [\; \mu_0 \lambda^2 \oint_{\partial \Omega_k}  \! \bm{j}(\bm{r}, t) \cdot d\bm{l} \; +  \Phi_k(t) \; ]\; ,
\end{equation} 
with $k$=1, ..., $N$-1. Here $\mu_0$ is the permeability of free space, $\lambda$ the London penetration depth, ${\bm{j}}({\bm{r}},t)$ the supercurrent density with $\bm{r}$ a spatial vector, the symbol  $\cdot$ means scalar product and $d \bm{l}$ is an integration line element. The integration contour $\partial \Omega_k$ is chosen as the inner boundary contour of hole $k$ as indicated in Fig. 1, integrated in counterclockwise direction.
In Eq. (3), $\Phi_k(t)$ is the time-dependent total magnetic flux that penetrates the hole area $\Omega_k$ (Fig. 1). The sum of the two terms in square brackets in Eq. (3) is called the fluxoid and is similar to London's fluxoid \cite{LON61} though here the fluxoid in Eq. (3)  is not quantised.\\
\par
The time-dependent total flux $\Phi_k(t)$ through the $k^{\text{th}}$ hole of the array is the sum of two parts, 
\begin{equation}
\Phi_k(t) = \Phi_k^{(a)} + \Phi_k^{(J)}(t) \; ,
\end{equation}
where $\Phi_k^{(a)}$ is the static applied flux through the hole $\Omega_k$,  {\it{i.e.}} $\Phi_k^{(a)} = B_a  A_k$ with $A_k$ the area of the hole $k$ and $B_a$ the applied magnetic inductance, $B_a = \mu_0 H_a$ where $H_a$ is the applied magnetic field. Here we restrict our investigation to the case where all hole areas $A_k$ are of the same size $A_h$, with $A_h = 2\, h \,w_h$ (Fig. 1). The flux $\Phi_k^{(J)}(t)$ in Eq. (4) is the flux spilled into the $k^{\text{th}}$ array hole and according to Biot-Savart's law, $\Phi_k^{(J)}(t)$ originates from the supercurrent density ${\bm{j}}({\bm{r}},t)$ that is flowing throughout the whole superconducting array.\\
\par
We will show in the following how the flux $\Phi_k^{(J)}(t)$ can be calculated and how the junction currents $I_k(t)$ in Eq. (1) can be expressed in terms of the differences of gauge invariant phase differences,$ \, \varphi_{k+1}(t)- \varphi_{k}(t)$, leading to a system of coupled first-order non-linear ordinary differential equations for the $\varphi_k(t)$'s, and an integro-differential equation for the stream function $g(x,y)$ (defined below), from which the time-averaged voltage of the array as a  function of the applied magnetic field, can be determined.

\subsection{\label{sec:level2}Biot-Savart's law, London's equation and the stream function equation for a parallel SQUID array with wide tracks, busbars and leads}

In order to calculate the magnetic flux $\Phi_k^{(J)}(t)$ of Eq. (4), we employ Biot-Savart's law and the second London equation. The SQUID array we wish to model is made out of a YBCO high-temperature superconducting thin-film of thickness $d$ = 0.125 $\mu$m and a London penetration depth (77 K) of approximately $\lambda$ = 0.3 $\mu$m. Because here $\lambda > d$, the supercurrent density through the thickness $d$ of the film is nearly homogeneous, independent of the $z$ direction, and therefore Biot-Savart's law in 2D can be applied \cite{BRA05,CLE05}. In this case, the magnetic field $H^{(J)}(x,y,t)$ in $z$ direction, produced by the supercurrent density $\bm{j}(x,y,t)$ flowing in the array of domain $\Omega$ (see Fig. 1), is given by
\begin{equation}
H^{(J)}(x,y,t)= 
\end{equation}
\[
\frac{d}{4 \pi} \int_{\Omega} \frac{j_x(x',y',t) \,(y-y') - j_y(x',y',t) \,(x-x')}{\sqrt{(x-x')^2+(y-y')^2 \,}^{\;3}} dx' dy' ,
\]
where $j_x$ and $j_y$ are the $x$ and $y$ components of the supercurrent density ${\bm{j}}$.
One can express $j_x$ and $j_y$ in Eq. (5) in terms of the stream function $g(x,y,t)$ which is defined as
\begin{equation}
j_x = \frac{1}{d} \frac{\partial g}{\partial y}  \quad \text{and} \quad j_y = - \frac{1}{d} \frac{\partial g}{\partial x} \,.
\end{equation}
This type of stream function approach has been used previously by Khapaev \cite{KHA01, KHA05, KHA10} and Brandt \cite{BRA05}.\\
\par
To be allowed to integrate Eq. (5) in 2D by parts, it is required to smoothen the functions in the integrand of Eq. (5) to generate continuously differentiable functions. We achieve this by analytically integrating  $(y-y')/\sqrt{(c-x')^2+(y-y')^2}^{\,3}$ and  $(x-x')/\sqrt{(c-x')^2+(y-y')^2}^{\, 3}$  over small intervals around their singularity points. This leads to 
\begin{equation}
H^{(J)}(x,y,t) = f_s(x,y,t) 
\end{equation}
\[
- \, \frac{1}{4 \pi} \int_{\Omega} Q_F(x,y,x',y') g(x',y',t) \; dx' dy' \, ,
\]
where the kernel $Q_F(x,y,x',y')$ is defined in Appendix B, and 
\begin{equation}
f_s(x,y,t) 
\end{equation}
\[
= \, \frac{1}{4 \pi} \oint_{\partial \Omega} \frac{g(x',y',t)}{\sqrt{(x-x')^2 + (y-y')^2}^{\;3} }\, 
 \left( {\begin{array}{cc}
   x-x' \\
   y-y' \\
  \end{array} } \right)
\cdot {\bm{n}} \, dl' \, .
\]
Here $dl'$ is the integration line element and ${\bm{n}}$ is a 2D normal vector in the $xy$ plane which is perpendicular on the domain boundary $\partial \Omega$, pointing outwards, away from the area $\Omega$. The contour $\partial \Omega$ includes the hole-boundaries of the array.\\

\par
Exploiting the stream function symmetry about the $x$ axis, {\it {i.e.}} $g(x',y',t) = g(x',-y',t)$, one can restrict the integration domain $\Omega$ in Eq. (7) and the contour integration domain $\partial \Omega$ in Eq. (8) to only the upper domains $\Omega^u$ and $\partial \Omega^u$ ($y' \geq 0$), respectively, where $\Omega = \Omega^u \, \cup \, \Omega^d$, with the superscript $u$ referring to the upper ($y' > 0$) and $d$ to the lower ($y' < 0$) domain (see Fig. 2). Thus one finds for Eq. (7)
\begin{equation}
H^{(J)}(x,y,t) = f_s(x,y,t)  
\end{equation}
\[
- \, \frac{1}{4 \pi} \int_{\Omega^u} Q(x,y,x',y') \, g(x',y',t) \, dx' dy' \, ,
\]
where the integration is now only over the upper domain $\Omega^u$ with a kernel $Q$ given by 
\begin{equation}
Q(x,y,x',y') = Q_F(x,y,x',y') + Q_F(x,y,x',- y') \; ,
\end{equation}
and
\begin{equation}
f_s(x,y,t) = f_s^u(x,y,t) + f_s^u(x,-y,t) \; .
\end{equation}
Here $f_s^u$ is defined by Eq. (8) but with a contour integration only over the upper contour $\partial \Omega^u$ of the domain $\Omega^u$ (see Fig. 2), where $\partial \Omega^u$ includes the upper boundary part of the holes. In Eq. (11), the second term on the right hand side, due to symmetry, accounts for the lower domain part.\\

\par

In order to obtain an equation to  calculate the stream function $g(x,y,t)$ we use the second London equation which has the form
\begin{equation}
\lambda^2 \;{\bm{\nabla}} \times {\bm{j}} = - {\bm{H}} \; ,
\end{equation}
where ${\bm{H}}$ is the total magnetic field. Again, because $\lambda > d$, and thus $\bm{j}(x,y,z) = \bm{j}(x,y)$, we can employ the 2D second London equation which from Eq. (12) becomes
\begin{equation}
\Lambda \;\Delta_{xy} \;g(x,y,t) = H(x,y,t) \; .
\end{equation}
Here in 2D, the magnetic field and the stream function are governed by the 2D screening length or Pearl length \cite{PEA64}  $\Lambda = \lambda^2 / d$. The operator $\Delta_{xy}$ is the 2D Laplace operator, $\partial^2 / \partial x^2 + \partial^2 / \partial y^2$, while $H(x,y,t)$ is the total magnetic field in $z$ direction. Because
\begin{equation}
H(x,y,t) = H_a + H^{(J)}(x,y,t) \; ,
\end{equation}
where $H_a$ is the applied magnetic field in $z$ direction, one finds, using Eqs. (8), (9), (11), (13) and (14), a 2D second-order linear Fredholm integro-differential equation for the stream function $g( x,y,t)$ of the form
\begin{equation}
\Lambda \, \Delta_{xy} \, g(x,y,t)  + \frac{1}{4 \pi} \int_{\Omega^u} Q(x,y,x',y') \, g(x',y',t) \, dx' dy' 
\end{equation}
\[
= \, H_a  +  f_s^u(x,y,t)  + f_s^u(x,-y,t) \;,
\]
with
\begin{equation}
f_s^u(x,y,t)  = 
\end{equation}
\[
\frac{1}{4 \pi} \oint_{\partial \Omega^u} \frac{g(x',y',t)}{\sqrt{(x-x')^2 + (y-y')^2}^{\;3}} \, 
 \left( {\begin{array}{cc}
   x-x' \\
   y-y' \\
  \end{array} } \right)
\cdot {\bm{n}} \, \,dl' \, .
\] 

\subsection{\label{sec:level2}Array boundary condition for the stream function intergo-differential equation}

In order to solve Eq. (15) one has to define boundary conditions for $g(x,y,t)$ and $\Delta_{xy} g(x,y,t)$ along the boundary $\partial \Omega^u$. 

Figure 2 shows the names given to different sections along the $\partial \Omega^u$ boundary. As the bias-current $I_b$ is injected downwards into the top current lead (Fig. 1), and because $j_y d = - \, \partial g / \partial x$, we choose for the current injection boundary condition $g(x,y,t) = I_b \; x / (2 c)$ for $(x,y) \, \epsilon \, \partial \Omega^{(T)}$. Because no current is crossing the boundaries $\partial \Omega^{(L)}$, $\partial \Omega^{(R)}$ and $\partial \Omega_k$ (inside holes), we find by using Eq. (6) the boundary conditions
$g(x,y,t) = - I_b/2$ for  $(x,y) \, \epsilon \, \partial \Omega^{(L)}$ and $g(x,y,t) = I_b/2$ for $(x,y) \, \epsilon \, \partial \Omega^{(R)}$, while for the hole-boundary conditions along $(x,y) \epsilon \, \partial \Omega_k$, one has $g(x,y,t) = \tilde g_k(t)$ where
\begin{equation}
\tilde g_k(t) = \sum_{j=1}^{k} I_j(t) \, - \, \frac{I_b}{2} \; ,
\end{equation}
Thus, the junction currents $I_1(t)$ and $I_N(t)$ at the ends of the array are $I_1(t) = \tilde{g}_1(t) + I_b / 2$ and $I_N(t) = -\tilde{g}_{N-1}(t) + I_b /2$, while between holes, the junction currents are $I_k(t) = \tilde{g}_k(t) - \tilde{g}_{k-1}(t)$.\\

With the above boundary conditions one can calculate $f_s(x,y,t)$ (Eqs. (11) and (16)) which can be written in the form
\begin{equation}
f_s(x,y,t) = P_0(x,y) \, I_b  +  \sum_{k=1}^{N-1}  \, P_k(x,y) \, \tilde g_k(t) \; .
\end{equation}
where both $P_0(x,y)$ and $P_k(x,y)$ are independent of time $t$ and
\begin{equation}
P_0(x,y) : = P_0^u(x,y) + P_0^u(x,-y) 
\end{equation}
\[
\quad \text{and} \quad P_k(x,y) : = P_k^u(x,y) + P_k^u(x,-y) \; .
\]
Here $P_0^u$ is the part of the integral in Eq. (16) along the contours $\partial \Omega^{(R)} \, \cup \, \partial \Omega^{(T)} \, \cup \, \partial \Omega^{(L)}$, while $P_k^u$ is the part of the integral along the upper half ($y \geq 0$) of the hole-contour $\partial \Omega_k$ (see Fig. 2).
We have calculated $P^u_0(x,y)$ and $P^u_k(x,y)$ analytically using Eq. (16) and the results are shown in Appendix B.\\ 

\begin{figure}[h]\begin{center}
\includegraphics[width=0.5\textwidth]{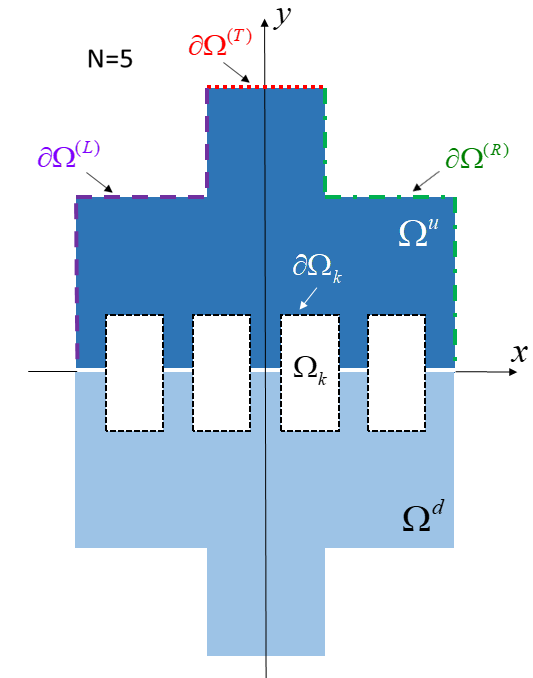}
\caption{The different boundary contour sections $\partial \Omega^{(L)}$,  $\partial \Omega^{(T)}$, $\partial \Omega^{(R)}$  along the upper domain $\Omega^u$ (shaded darker) and the hole domains $\Omega_k$ and contours $\partial \Omega_k$ for $k=1, ... , N-1$, of the holes between domain $ \Omega^u$ and $\Omega^d$.} 
\label{default}
\end{center}
\end{figure}

\par

\subsection{\label{sec:level2}Magnetic flux in array holes}

The total magnetic flux $\Phi_k(t)$ (Eq. (4)) in hole $k$, required in Eq. (3), becomes with Eq. (14),
\begin{equation}
\Phi_k (t) = \mu_0 \int_{\Omega_k}  ( \, H_a + H^{(J)}(x,y,t) \; )\, dx dy \; ,
\end{equation}
where the integration is over the hole domain $\Omega_k$ (see Fig. 2).\\

Using Eqs. (9) and (18), one derives
\begin{equation}
\frac{\Phi_k(t)}{\mu_0} = H_a A_h + L_{k 0} \, I_b + \sum_{j=1}^{N-1} L_{kj} \, \tilde g_j(t) 
\end{equation}
\[
- \, \frac{1}{4 \pi} \int_{\Omega^u}  \tilde Q_k(x',y') \, g(x',y',t) \, dx' dy' \; ,
\] \\
where $\mu_0 H_a A_h$ is the applied magnetic flux that penetrates each hole of area $A_h$, and
\begin{equation}
L_{kj}: = \int_{\Omega_k} P_j(x,y) \, dx dy = 2 \int_{\Omega_k} P_j^u(x,y) \,dx dy \; ,
\end{equation} 
with $j = 0,1, ... N-1$. Furthermore, in Eq. (21), $\tilde{Q}_k(x',y')$ is defined as
\begin{equation} 
\tilde Q_k(x',y') : = \int_{\Omega_k}  Q(x,y,x',y') \, dx dy \, ,
\end{equation}
where $Q(x,y,x',y')$ is given by Eq. (10) with $(x',y') \, \epsilon \, \Omega^u$. $L_{kj}$ in Eq. (22) as well as $\tilde Q_k(x',y')$ in Eq. (23) can be calculated analytically. But here, calculations were simply performed numerically for each hole $k$, since these quantities are solely of geometrical nature and time-independent, and thus, have to be calculated only once.\\

\par
\subsection{\label{sec:level2}Conversion to algebraic equations and vectorization}

In order to calculate numerically the stream function $g(x,y,t)$ with Eq. (15) for $(x,y) \, \epsilon \, \Omega^u$, one has to convert Eq. (15) into an algebraic equation. To do so, we choose a sufficiently fine square grid on $\Omega^u$ (Fig. 2) and discretise the spatial vectors $\bm{r} = (x,y)$ to ${\bm{r}_n} = (x_n,y_n)$ located at the centre of each small square grid element of size $w = (\Delta x )^2$, where $\Delta x$ is the square grid spacing and the index $n$ counts the grid elements, $n = 1, ..., N_g$, where $N_g$ is the total number of grid elements in the domain $\Omega^u$. The integro-differential equation Eq. (15) for $g(x,y,t)$ then becomes
\begin{equation}
\sum_{m \, \epsilon \, \Omega^u} \left[ \, \Lambda  \,\Delta_{nm} \, + \frac{w}{4 \pi} Q_{nm} \, \right] \, g_m(t) 
\end{equation}
\[
= \, H_a + P_{0n} I_b + \sum_{k=1}^{N-1} P_{nk}  \, \tilde g_k(t) \; ,
\]
for all $n \, \epsilon \, \Omega^u$. Here
\begin{equation}
g_n(t) : =g({\bm{r}}_n,t) \, , \quad Q_{nm} : = Q ({\bm{r}}_n, {\bm{r}}_m), 
\end{equation}
\[
\quad P_{0n} : = P_0(\bm{r}_n)  \quad \text{and} \quad P_{nk} : = P_k ({\bm{r}}_n) \; .
\]
\\
\par
In vector notation, the relationship in Eq. (17) between the boundary values $\tilde{g}_k(t)$ and the junction currents ${I_k}(t)$ and bias injection current $I_b$ becomes
\begin{equation}
\tilde{\bm{g}}(t) = \mathcal{T} \circ \tilde{\bm{I}}(t) - \frac{I_b}{2} \, \bm{1}_{N-1} \; ,
\end{equation}
where $\tilde{\bm{g}}$ is the $N-1$ dimensional stream function vector, $\tilde{\bm{g}} = (\tilde{g}_1, ... , \tilde{g}_{N-1})$ for the holes. In Eq. (26) the symbol ${\circ}$ means multiplication of a matrix with a vector (and also later matrix multiplication) and the above $(N-1) \times (N-1)$ matrix $\mathcal{T}$ is defined as $(\mathcal{T})_{kj} = 1$ if $k \geq j$ and zero otherwise, and the junction current vector  $\tilde{\bm{I}}(t)$ is an $N-1$ (not $N$) vector,  $\tilde{\bm{I}}(t) : = (I_1(t), ... , I_{N-1}(t))$.  The vector $\bm{1}_{N-1}$ is of dimension $N-1$ where $\bm{1}_{N-1} : = (1, ... , 1)$.
By using Eq. (26) above, Eq. (24) can conveniently be written in vector notation, and by performing a matrix inversion one obtains the time-dependent stream function vector $\bm{g}(t)$ as

\begin{widetext}
\begin{equation}
{\bm{g}}(t) = \left( \, \Lambda  \mathlarger{{\mathcal{D}}}  + \frac{w}{4 \pi} \, \mathlarger{\mathcal{Q}} \, \right)^{-1} \circ
\left[ \, H_a \bm{1}_{N_g} +  \bm{P}_0 \,\, I_b \, + 
\mathcal{P} \circ \mathcal{T} \circ \tilde{\bm{I}}(t) + [  \bm{P}_0 - \frac{1}{2} \mathcal{P} \circ \bm{1}_{N-1} ] \, I_b  - \Lambda \, \bm{d}_{\Delta}(t)  \right] \; .
\end{equation}
\end{widetext}

The stream function vector, $\bm{g}(t)  : = (g_1(t), ... , g_{N_g}(t))$, represents the stream function $g(x,y,t)$ at all grid point elements in $\Omega^u$. The matrix $\mathlarger{\mathcal{D}}$ is an $N_g \times N_g$ matrix corresponsing to the Laplace operator in Eq. (24), where $(\mathlarger{\mathcal{D}})_{nm} : = \,[ \,-4 \, \delta_{nm} + \delta^{Nb}_{nm} \, ] \, / w$ with $w = (\Delta x)^2$ and $\delta^{Nb}_{n,m} = 1$ if $\bm{r}_m$ is a nearest neighbour of $\bm{r}_n$ and zero otherwise. The matrix $\mathlarger{\mathcal{Q}}$ is also an $N_g \times N_g$ matrix, defined as $(\mathlarger{\mathcal{Q}})_{nm} : = Q(\bm{r}_n,\bm{r}_m)$. The symbol $\bm{1}_{N_g}$ in Eq. (27) is the vector $\bm{1}_{N_g} = (1,1,...,1)$ of dimension $N_g$ and $\bm{P}_0 : = (P_1, ... , P_{N_g})$, defined in Eq. (19) and Appendix B. The symbol $\mathlarger{\mathcal{P}}$ is an $N_g \times
(N-1)$ matrix with $(\mathlarger{\mathcal{P}})_{nk} : = P_k(\bm{r}_n)$. The components of the time-dependent $N_g$-dimensional vector $\bm{d}_{\Delta}(t)$ in Eq. (27) originate from the part of the Laplace operator which operates on the domain boundary $\partial \Omega^u$ (which includes the holes) and the boundary along junctions. Most of the components of $\bm{d}_{\Delta}(t)$ are zero, but for grid elements adjacent to boundaries, the components are $8/3 \, \tilde{g}^{(L)}/ w$, $8/3 \, \tilde{g}^{(T)}/w$, $8/3 \, \tilde{g}^{(R)}/w$, and $8/3 \, \tilde{g}_k(t)/w$ (along $\partial \Omega_k$) (Fig. 2). For grid elements adjacent to junctions, the corresponding components of  $\bm{d}_{\Delta}(t)$ vary linear with distance along the junctions. The time dependence of  $\bm{d}_{\Delta}(t)$ therefore arises from the time-dependent hole-boundary conditions (Eq. (17)). The above factor of $8/3$ results from using a Laplacian for a non-equidistant grid (Eq. (C1) in Appendix C). Care has to be taken at corner grid elements. Note that in order to calculate $\bm{g}(t)$ using Eq. (27), a very large $N_g \times N_g$ matrix, $\Lambda \mathcal{D} + \frac{w}{4 \pi} \mathcal{Q} $, has to be inverted. It is important to note that since this matrix is time-independent, this large matrix inversion has to be performed only once at the beginning of a computation.\\

\par

In vector notation, using Eqs. (21) and (26), the flux vector $\bm{\Phi}(t) = (\Phi_1(t), ... , \Phi_{N-1}(t))$ of the magnetic flux inside array holes, takes the form
\begin{equation}
\frac{\bm{\Phi}(t)}{\mu_0} = H_a A _h\, \bm{1}_{N-1} + ( \, { \bm{L}}_0 \, - \frac{1}{2}  \,{{\mathcal{L}}} \circ \bm{1}_{N-1}\, )\, I_b 
\end{equation}
\[
+ \, {\mathcal{L}}  \circ \mathcal{T}  \circ \tilde{\bm{I}}(t) \,-\, \frac{w}{4 \pi} \tilde{\mathcal{Q}} \circ \bm{g}(t) \; ,
\] \\
where ${\bm{L}}_0 : = (L_{10}, ... , L_{N-1,0})$ with $L_{k 0}$ given by Eq. (22), and $\mathcal{L}$ is an  $(N-1) \times (N-1)$ matrix defined as $({\mathcal{L}})_{kj} : = L_{kj}$ ($j \geq 1$) where $L_{kj}$ is again given by Eq. (22). In Eq. (28) $\tilde{\mathcal{Q}}$ is an $(N-1) \times N_g$ matrix given by $(\tilde{\mathcal{Q}})_{kj} : = \tilde{Q}_k(\bm{r}_j)$ where $\tilde{Q}_k(\bm{r}_j)$ is defined by Eq. (23) .\\

\par

Furthermore, we rewrite the Ginzburg-Landau equation Eq. (3) in vector notation of the form
\begin{equation}
\mathcal{N} \circ \bm{\varphi}(t) = \frac{2 \pi}{\Phi_0} ( \, \mu_0 \, \Lambda \, \bm{K}(t) + \bm{\Phi}(t) \,) \; ,
\end{equation}
where $\mathcal{N}$ is an $(N-1) \times N$ matrix defined as $(\mathcal{N})_{kj} = - 1$ for $k = j$ and  $(\mathcal{N})_{kj} = 1$ for $k = j - 1$ and zero otherwise. The phase difference vector $\bm{\varphi}(t)$ is $N$ dimensional, $\bm{\varphi}(t): = (\varphi_1(t), ... , \varphi_N(t))$. Using Eqs. (3) and (6), the components $K_k(t)$ of the $(N-1)$ dimensional vector $\bm{K}(t) : = (K_1(t), ... , K_{N-1}(t))$ are 
\begin{equation}
K_k(t)= d \, \oint_{\partial \Omega_k} \bm{j}(t) \cdot d\bm{l} = \oint_{\partial \Omega_k} (\, \frac{\partial g(t)}{\partial y} \, dx - \frac{\partial g(t)}{\partial x} dy  \,) \; ,
\end{equation}
where the contour integration is counterclockwise along the boundary $\partial \Omega_k$ of the hole $k$ (see Figs. 1 or 2).\

\par
\subsection{\label{sec:level2}System of coupled differential equations for the Josephson phase differences for a parallel SQUID array with wide thin-film structure}

Using Eq. (28) and (29) and eliminating the flux vector $\bm{\Phi}(t)$, one derives an equation for the junction current vector $\tilde{\bm{I}}(t)$ as a function of the phase difference vector $\bm{\varphi}(t)$ and the stream function vector $\bm{g}(t)$ of the form

\begin{widetext}
\begin{equation}
\tilde{\bm{I}}(t) = \left( \mathcal{L} \circ \mathcal{T} \right)^{-1}\, \left[ \;\frac{\Phi_0}{2 \pi \, \mu_0} \, \mathcal{N} \circ \bm{\varphi} (t)- \Lambda \, \bm{K}(t) - 
H_a \, A_h\, \bm{1}_{N-1} - \left( \, \bm{L}_0 - \frac{1}{2} \,\mathcal{L} \circ \bm{1}_{N-1} \, \right)  \, I_b  
                      + \frac{w}{4 \pi} \tilde{\mathcal{Q}} \circ \bm{g}(t) \, \right] \; .
\end{equation}      
\end{widetext}         

Note that $\tilde{\bm{I}}(t)$ is an $(N-1)$ vector and its components do not contain the junction current $I_N(t)$ across the last JJ. But, since $I_N(t) = I_b - \sum_{k=1}^{N-1} I_k(t)$, the current $I_N(t)$ is well defined and thus, Eq. (31), together with Eqs. (1) and (2) define a complete set  of coupled first-order differential equations for the gauge invariant phase differences $\varphi_k(t)$ with $k=1, ..., N$.\\

It is convenient to define $I_c$ as the average junction critical current, $I_c = \sum_{k=1}^{N} I_{ck} \, / \,N$ and $R$ as the average junction resistance, $R = \sum_{k=1}^{N} R_k \, /  \, N$. Note that $R$ is not the total array resistance. Then, Eq. (1) combined with Eq. (2) can be put into the form
\begin{equation}
\frac{ d \, \varphi_k(\tau) }{ d \tau} = \, \xi_k \,  \left( \, - \eta_k \, \sin \varphi_k(\tau) \, + \frac{I_k(\tau)}{I_c}
\, + \frac{I_c^{Noise}(\tau)}{I_c} \, \right)  \; ,
\end{equation}
where $\tau$ is the reduced time in dimensionless units,
\begin{equation}
\tau = \frac{2 \pi}{\Phi_0} \, R \, I_c \; t  \; ,
\end{equation}
and $\xi_k = R_k/R$ and $\eta_k = I_{ck}/I_c$, and thus $\xi_k$ ($\eta_k$) is a measure of deviation of $R_k$ ($I_{ck}$) from $R$ ($I_c$). The standard deviations for $R_k$ and $I_{ck}$ in YBCO thin-film SQUID arrays can be as large as $0.3$\, \cite{MIT19, LAM14}.\\

At this point it is convenient to write Eq. (32) in vector notation and combining it with Eq. (31) which results in
\begin{widetext}
\begin{equation}
\bm{\dot \varphi}(\tau) = \bm{S}(\tau) \, + \, \widehat{\mathcal{{\xi}}} \circ \left( \mathcal{L} \circ \mathcal{T} \right)^{-1}
\end{equation}
\[
\circ \, \left[ \; \frac{ \Phi_0}{2 \pi \, \mu_0}  \, \mathcal{N} \circ \bm{\varphi}(\tau) - \Lambda \, \bm{K}(\tau) + \frac{w}{4 \pi} \tilde{\mathcal{Q}} \circ \bm{g}(\tau) 
-  \left( \, \bm{L}_0 - \frac{1}{2} \, \mathcal{L} \circ \bm{1}_{N-1} \, \right)  I_b  \,- H_a \, A_h \,  \bm{1}_{N-1} \, \right] \, + \widehat{\mathcal{\xi}}\, \circ \, \tilde{\bm{I}}^{Noise}(\tau) \; .
\]
\end{widetext}
Here the components of the vector $\bm{\dot \varphi}(\tau)$ are $d \varphi_k(\tau) / d \tau$ and the components of the vector $\bm{S}(\tau)$ are $-\xi_k \, \eta_k \sin \varphi_k(\tau)$, where $k=1, ... , N-1$. The matrix $\widehat{\mathcal{\xi}}$ is an $(N-1) \times (N-1)$ diagonal matrix with diagonal elements $\xi_k / I_c$, where $k=1, ... ,N-1$, and $\tilde{\bm{I}}^{Noise}(\tau) : = (I_1^{Noise}(\tau), ... , I_{N-1}^{Noise}(\tau))$. Please note that $\bm{K}(\tau)$ and $\bm{g}(\tau)$ are both functions of time $\tau$ since the supercurrent density distribution in a SQUID array undergoes periodic changes with time.
\\ 
\par
To make the set of coupled differential equations for $\varphi_k(\tau)$ of Eq. (34) complete, one has to add an equation for $d \varphi_N(\tau) / d \tau$ using Eqs. (1) and (2), which gives, because of $I_b = \sum_{k=1}^N I_k(\tau)$,

\begin{widetext}
\begin{equation}
\frac{d \varphi_N(\tau)}{d \tau} = \xi_N  \left( -  \eta_N \sin \varphi_N(\tau) + \left[ I_b - \sum_{k=1}^{N-1}I_k(\tau) \right] / \, I_c  + I_N^{Noise}(\tau) / \, I_c \,  \right)\; .
\end{equation}
\end{widetext}

Equation (34) together with Eq. (35) forms a complete set of coupled first-order differential equations for all the $\varphi_k(\tau)$'s.  To solve this set of differential equations, initial conditions for all $\varphi_k(\tau=0)$ have to be chosen. This is done by starting with equal junction currents $I_k(\tau=0) = I_b / N$ and using Eqs. (26) - (28) and (30) to calculate the $\varphi_k(\tau=0)$'s from $\mathcal{N} \circ \bm{\varphi}(\tau)$ of Eq. (29), and setting $\varphi_1(\tau =0)$ = 0.\\
\par
Please note that Eqs. (34) and (35) together with Eq. (26), (27) and (31) are the key equations of this paper. 
The set of equations Eqs.(34) and (35) were solved numerically using the $4^{\text{th}}$ order Runge-Kutta method. After each time step, chosen as $\Delta \tau = 0.1$, the $\varphi_k$'s change and thus the Josephson currents $I_k(\tau)$ change slightly (Eq. (31)), which then slightly changes the boundary condition $\tilde {\bm{g}}(\tau)$ (Eq. (26)). Thus, after each time step $\Delta \tau$ an updated stream function $\bm{g}(\tau+\Delta \tau)$ (Eq. (27)) has to be calculated, resulting in updated $\bm{g}$ and $\bm{K}$ vectors in Eq. (34). \\
Details about how the noise currents $\tilde{\bm{I}}^{Noise}(\tau)$ in Eq. (34) and $I_N^{Noise}(\tau)$ in Eq. (35) were treated numerically are outlined in Appendix D.
\par
\subsection{\label{sec:level2}Time-averaged voltage}
The time-averaged normalised voltage, $V / (R I_c)$, between the leads of a parallel SQUID array, is given by time averaging the right-hand side of the Josephson equation Eq. (2) which results in
\begin{equation}
\frac{V}{R \, I_c} = \lim_{\tau \rightarrow \infty} \, \frac{1}{\tau} \, \left[ \;\varphi_k(\tau + \tau_0) - \varphi_k(\tau_0) \; \right] \; ,
\end{equation}
where $k$ can be any $k \, \epsilon \, \{1, ... , N\}$. One has to choose $\tau_0$ large enough such that numerical self-adjustment for the initial $\varphi_k$'s has occurred and $\tau$ has to be taken sufficiently large. \\

In the case where Johnson noise can be neglected, one finds
\begin{equation}
\frac{V}{R \, I_c} = \frac{2 \pi}{\tau_p} \; ,
\end{equation}
where $\tau_p$ is the period of oscillations of the $\varphi_k(\tau)$'s. In the case of non-negligible Johnson noise, Eq. (37) cannot be used but one can reduce the statistical error in $V / (R I_c)$ by averaging over all the $N$ phase differences, {\it{i.e.}}
\begin{equation}
\frac{V}{R I_c} \simeq \frac{1}{\tau} \; \frac{1}{N}  \sum_{k=1}^{N} \, [ \, \varphi_k(\tau + \tau_0) - \varphi_k(\tau_0) \, ] \; ,
\end{equation}
where $\tau$ has to be chosen sufficiently large. 

\par
\subsection{\label{sec:level2}Effective areas of SQUID array holes}

The wide tracks, busbars and leads focus magnetic flux into the array holes. In addition, the Meissner shielding current crowding near the holes, enhance the $\oint \bm{j} \cdot \bm{dl}$ term. An effective area, $A_k^{eff}$ can be defined for each array hole $k$ via the fluxoid it contains as 
\begin{equation}
A_k^{eff} = \lim_{I_c, I_b \rightarrow 0} \frac{\mu_0 \lambda^2 \oint_{\partial \Omega_k} \bm{j} \cdot d \bm{l}  + \Phi_k }{B_a} \; .
\end{equation}
$A_k^{eff}$ can also be defined in the limit of $B_a \rightarrow \infty$ instead of $I_c$ and $I_b \rightarrow 0$. From the definition in Eq. (39), employing Eqs. (28) and (30), it follows that $A_k^{eff}$ in vector notation, $\bm{A}^{eff}$, normalised to the hole area $A_h$, is
\begin{equation}
\frac{\bm{A}^{eff}}{A_h} = \bm{1}_{N-1} + \frac{\mu_0}{B_a A_h} \left( \, \Lambda \, \bm{K}_0 - \frac{w}{4 \pi} \tilde{\mathcal{Q}} \circ \bm{g}_0  \, \right) \; ,
\end{equation}
where the subscript $0$ in $\bm{K}_0$ and $\bm{g}_0$ means that $\bm{g}$ in Eq. (27) and $\bm{K}$ in Eq. (30) are evaluated with the boundary condition $g(x,y)$ = 0 along $\partial \Omega^u$, since $I_c=I_b = 0$.

\section{Device fabrication and experiment}

 Parallel SQUID arrays were fabricated lithographically by growing thin-films of YBCO on 1 cm$^2$ MgO substrates. Steps were etched into the MgO surface using a well established technique based on argon milling \cite{FOL99,MIT10}. During YBCO thin-film growth by e-beam evaporation, a long grain boundary forms at the top of the edge of the MgO step, creating a long JJ. Films were then lithographically patterned into parallel SQUID arrays \cite{MIT19}. The width of junctions and junction tracks is $w_J=2 \, \mu$m and the width of holes $w_h=4\mu$m with half-height $h=4 \,\mu$m (see Fig. 1). For our $N = 11$ array, which is the array discussed in detail in this paper, the thickness of the thin-film is $d$ = 0.125 $\mu$m, the width of the busbar $b-h=8 \, \mu$m and the bias-lead half-width $c=4 \, \mu$m. For the calculation the length of the bias-lead was chosen sufficiently long as $l$ = 24 $\mu$m (see Fig. 1). A micrograph of the $N = 11$ device is shown as an inset in Fig. 3(a).\\
 \par
To measure the $V(B_a)$ response, the array was placed on a measurement probe, that generated a perpendicular applied induction $B_a$ (magnetic field $H_a$), and then dipped into a dewar of liquid nitrogen and zero field cooled down to a temperature of 77 K. To screen out  the earth's magnetic field, the dewar was surrounded by five layers of mu-metal shielding. A bias-current $I_b$, with a value which optimised the SQUID array response (maximised transfer function $d V / d B_a$), was injected into the top bias-lead and the voltage $V$ at different $B_a$ was measured using the standard four-terminal method.\\
\par

\section{Results and discussion}

In the following we discuss in detail the calculated and experimental results for a parallel SQUID array with $N = 11$ junctions with 8 $\mu$m wide fluxoid focussing busbars and bias-leads. A micrograph of this device is shown as an inset in Fig. 3(a). The parameters used in the calculation are the array bias-current $I_b$ = 200 $\mu$A (given by experiment), the average critical current of a  junction $I_c$ = 24 $\mu$A and the London penetration depth $\lambda$ = 0.33 $\mu$m.
Further below we will discuss how the values for $I_c$ and $\lambda$ were determined. We found that choosing the square grid spacing as $\Delta x$ = 0.5 $\mu$m, or even $\Delta x$ = 1 $\mu$m, was numerically sufficient and therefore the number of grid points $N_g$ that lie in the array domain $\Omega^u$ is $N_g$ = 3104, or 776.
\\
\par

\begin{figure}[h]
\begin{center}
\includegraphics[width=0.56\textwidth]{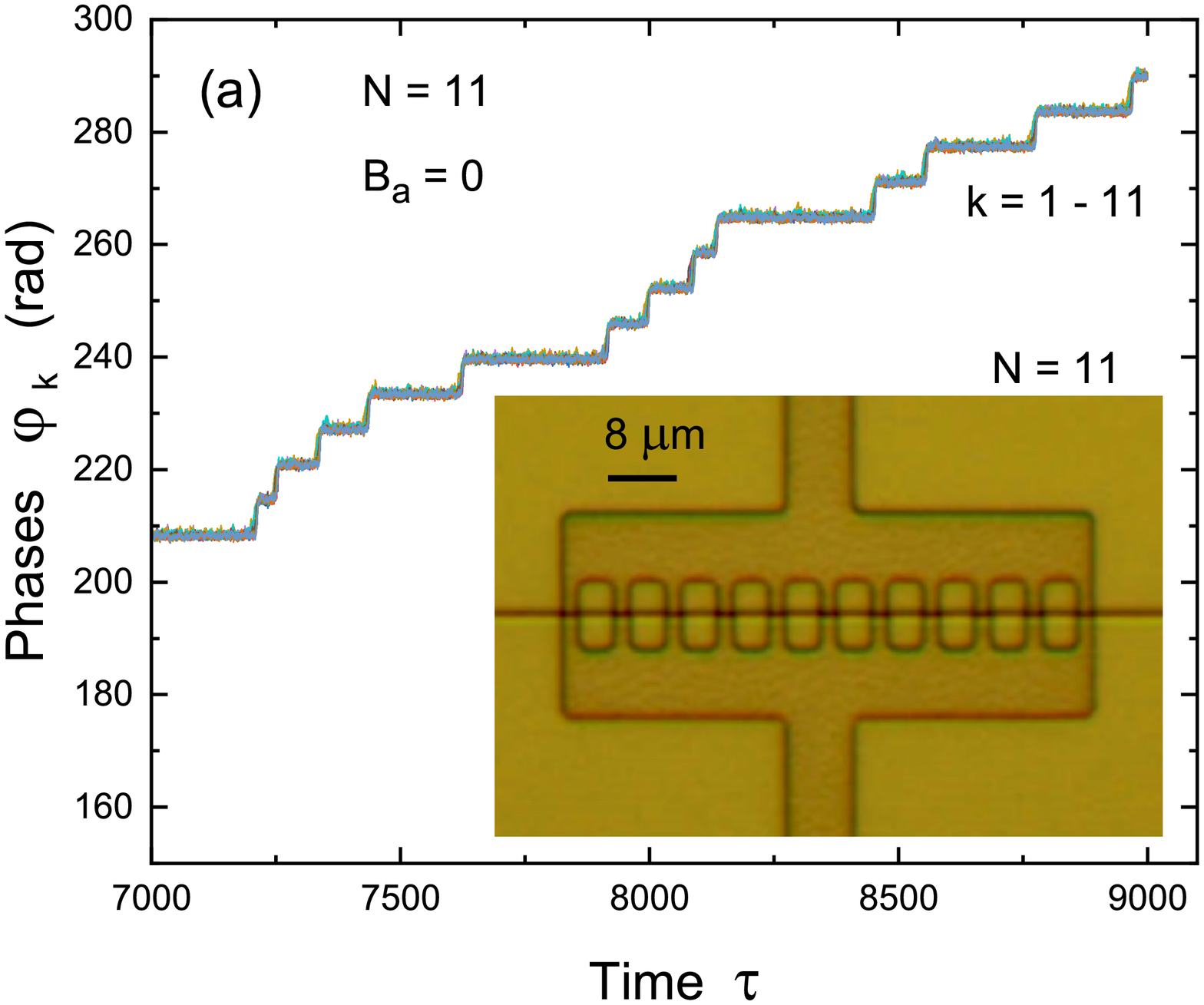}
\label{default}
\includegraphics[width=0.56\textwidth]{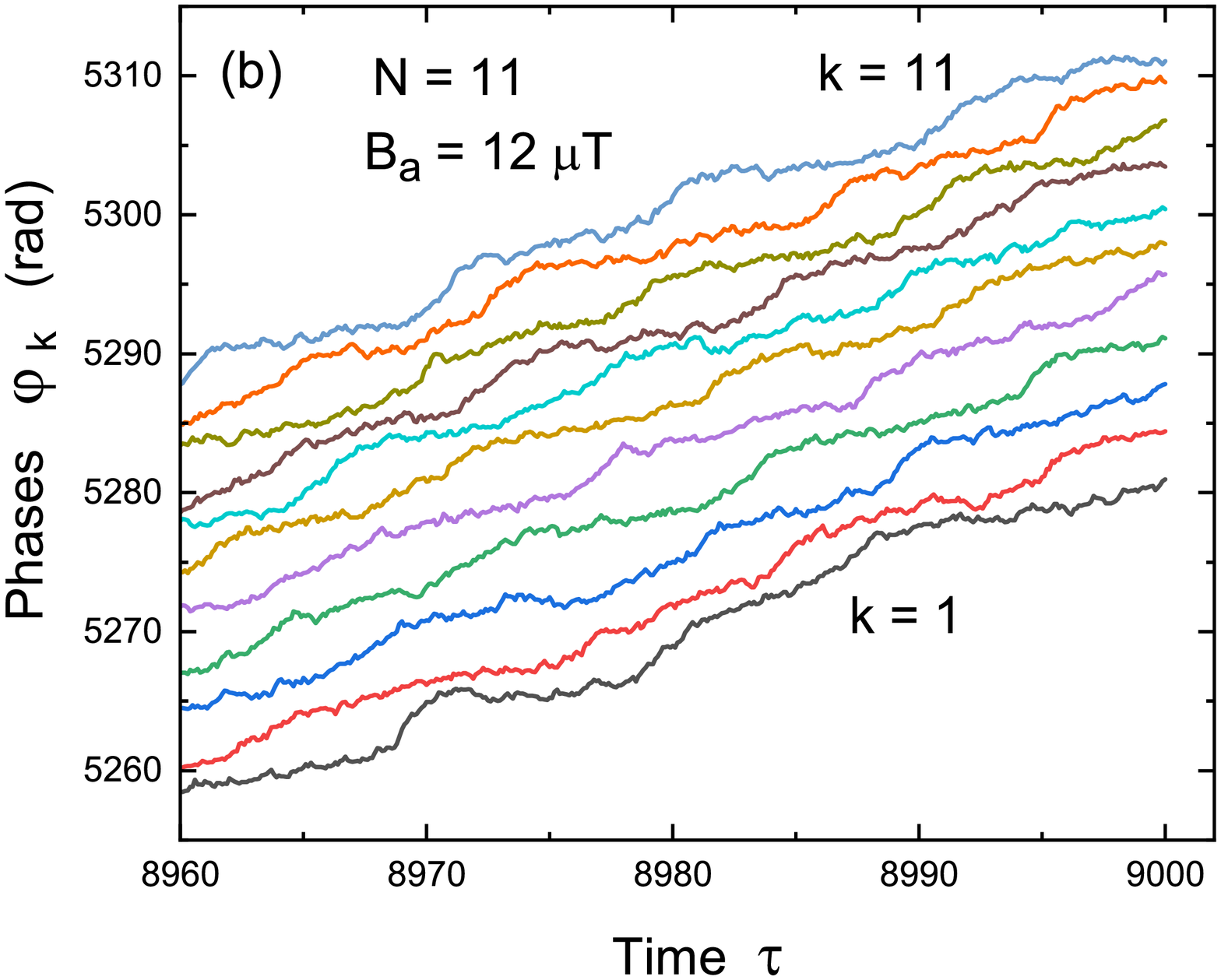}
\label{default}
\end{center}
\caption{The inset in (a) shows a micrograph of our parallel SQUID array with $N = 11$ junctions where the horizontal line indicates the step edge across which JJs have formed. (a) displays the time-evolution of the phase differences $\varphi_k$ for $k=1$ to 11, calculated using Eqs. (34) and (35) together with Eq. (27) for $B_a = 0$. (b) displays the time-evolution of the phase differences $\varphi_k(\tau)$ for $B_a$ = 12 $\mu$T.}
\end{figure}

\begin{figure}[h]
\begin{center}
\includegraphics[width=0.56\textwidth]{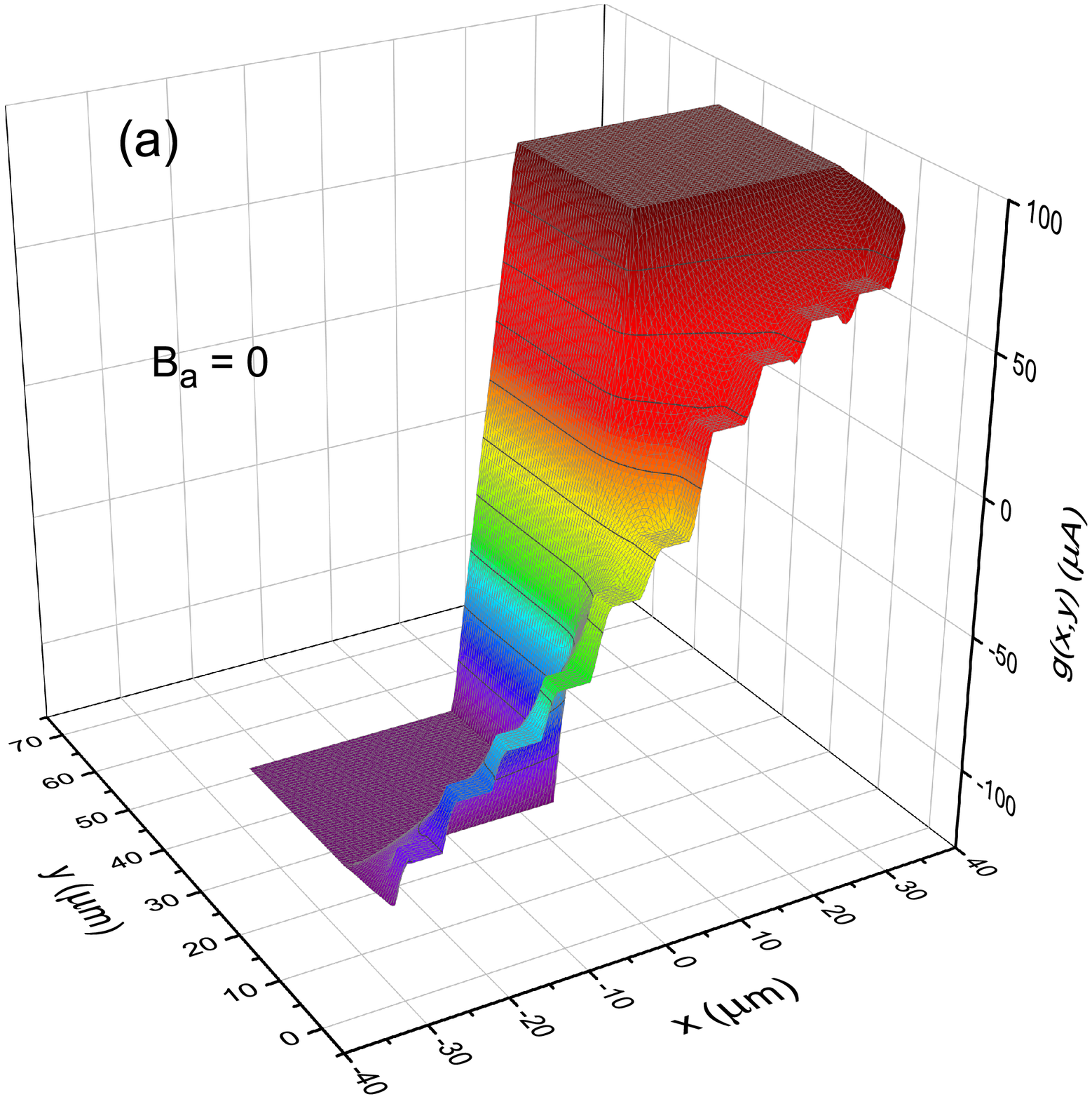}
\label{default}
\includegraphics[width=0.56\textwidth]{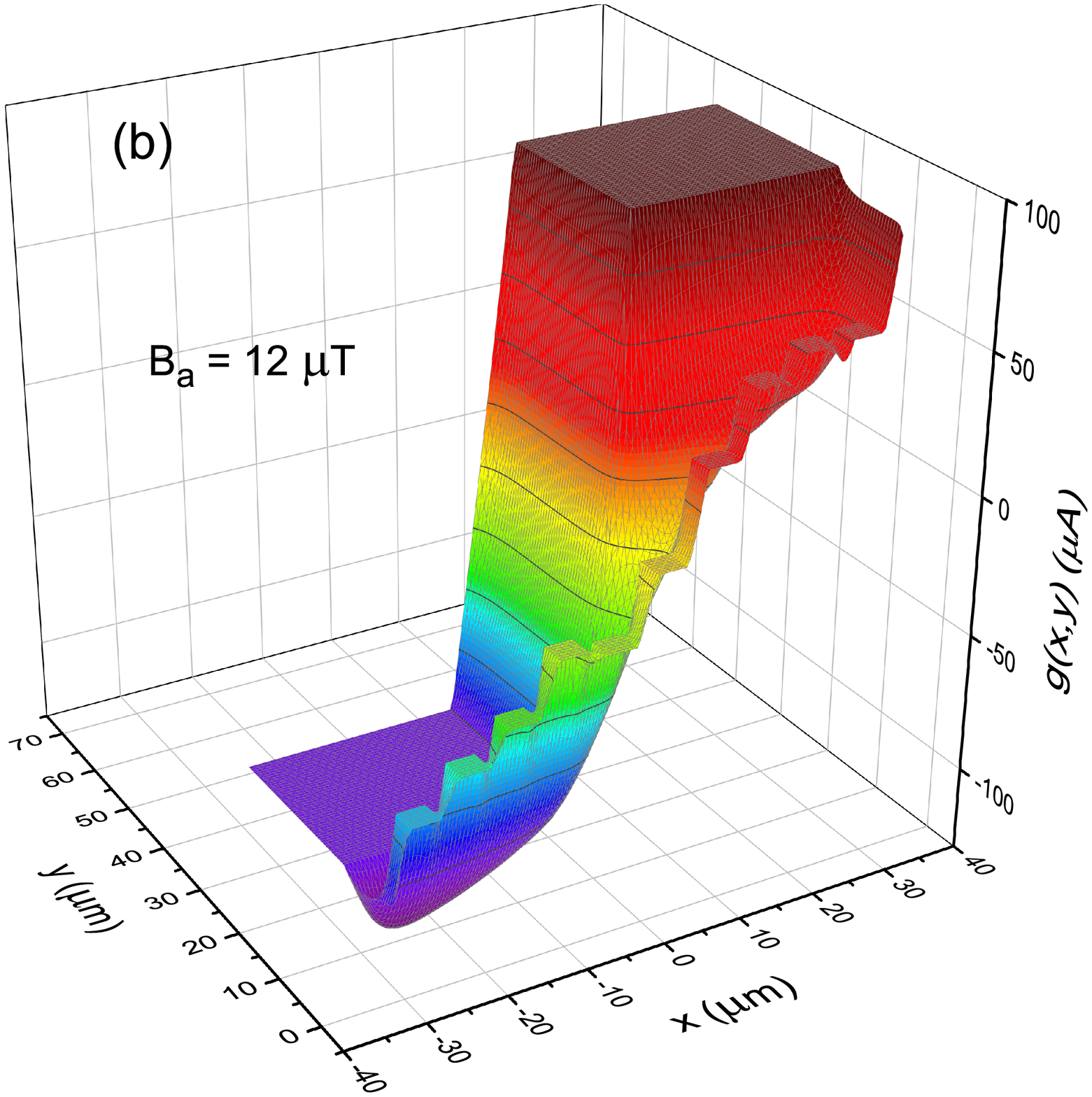}
\label{default}
\caption{Stream function $g(x,y)$ in the upper part ($y > 0$) of the $N$ = 11 parallel SQUID array at $\tau$ = 9000 for  (a) $B_a =0$ and (b) $B_a$ = 12 $\mu$T.}
\end{center}
\end{figure}

Figures 3(a) and (b) show the calculated time-evolution of the $N$ = 11 phase differences $\varphi_k(\tau)$ (from Eqs. (34) and (35)) for perpendicular applied magnetic inductions $B_a = 0$ and $B_a$ = 12 $\mu$T, repectively. For $B_a$ = 0, the fluxoid in each hole is close to zero, while at $B_a$ = 12 $\mu$T, the fluxoid in each hole is about half a flux quantum, $\Phi_0 / 2$. Due to thermally activated phase slippages, caused by an effective Johnson noise strength $\Gamma$ = 0.135 (Eq. (D1)) at $T = 77$K, the time-evolution of the $\varphi_k(\tau)$ in Figs. 3(a) and (b) is somewhat erratic. For $B_a = 0$ coherent phase slippages by $2 \pi$  occur quite suddenly, while in the $B_a$= 12 $\mu$T case, $2 \pi$ increments appear more sinusoidal. Furthermore, in the $B_a$ = 12 $\mu$T case the $\varphi_k(\tau)$ are incrementally shifted upwards by about $\pi$ with increasing $k$.
\\
\par
Figures 4(a) and (b), using Eq. (27), show the stream function $g(x,y)$ for the upper domain $\Omega^u$ at time $\tau$ = 9000 (Fig. 3) for $B_a = 0$ and $B_a$ = 12 $\mu$T, respectively. As can be seen, the stream function values along the left and right boundaries, $\partial \Omega^{(L)}$ and $\partial \Omega^{(R)}$, are $\mp \, I_b / 2$ = $\mp \, $100 $\mu$A, while the step-like structure of $g(x,y)$ along the 10 holes corresponds to the stream function boundary values $\tilde g_k$ in the holes which vary with time $\tau$. 
\\
\par

\begin{figure}[h]\begin{center}
\includegraphics[width=0.53\textwidth]{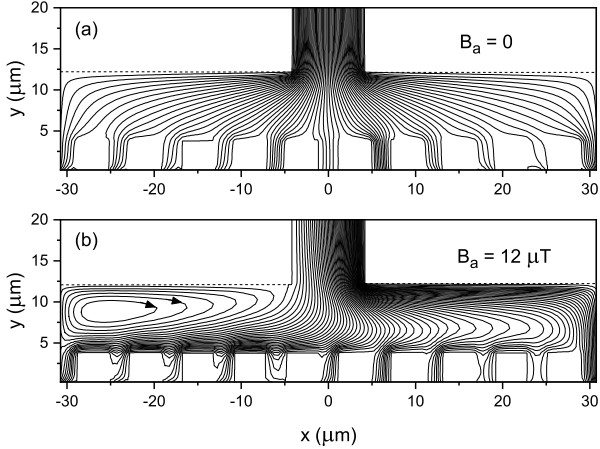}
\caption{Current stream lines in the upper domain $\Omega^u$ of the $N$ = 11 parallel SQUID array at time $\tau = 9000$ for (a) $B_a = 0$ and (b) $B_a$ = 12 $\mu$T.}
\label{default}
\end{center}
\end{figure}

Figures 5(a) and (b) display the current stream lines at $B_a =0$ and $B_a$ = 12 $\mu$T, obtained from the contour lines of $g(x,y)$ of Figs. 4(a) and (b). Only the upper domain $\Omega^u$ is shown because of the symmetry about the $x$ axis where $g(x,y) = g(x,-y)$ and thus from Eq. (6), $j_x(x,y) = - j_x(x,-y)$ and $j_y(x,y) = j_y(x,-y)$. As can be seen, for $B_a = 0$, the current fans out from the  bias-lead to the junctions where the currents through individual junctions vary with time. Current crowding is visible left and right of the bias-lead and in particular at the corners between bias-lead and busbar due to Meissner shielding. For $B_a$ = 12 $\mu$T, circulating Meissner shielding currents are visible in the left part of the busbar. Strong current crowding now only occurs on the right side of the bias-lead and the right corner between bias-lead and busbar. In addition, strong current crowding is visible at the top of holes due to the Meissner shielding current circulating in the busbar.
\\
\par

\begin{figure}[h]
\begin{center}
\includegraphics[width=0.58\textwidth]{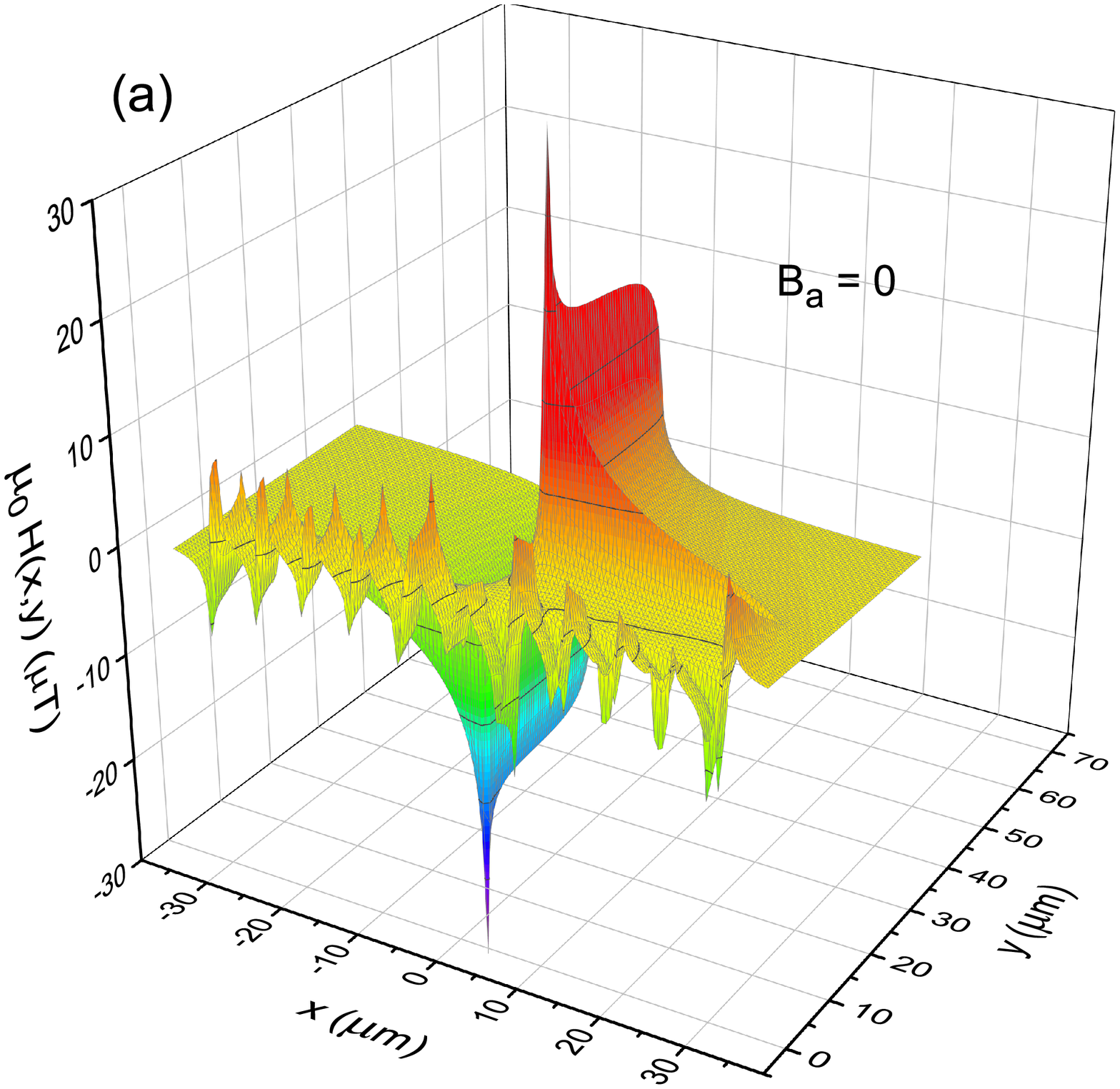}
\label{default}
\includegraphics[width=0.58\textwidth]{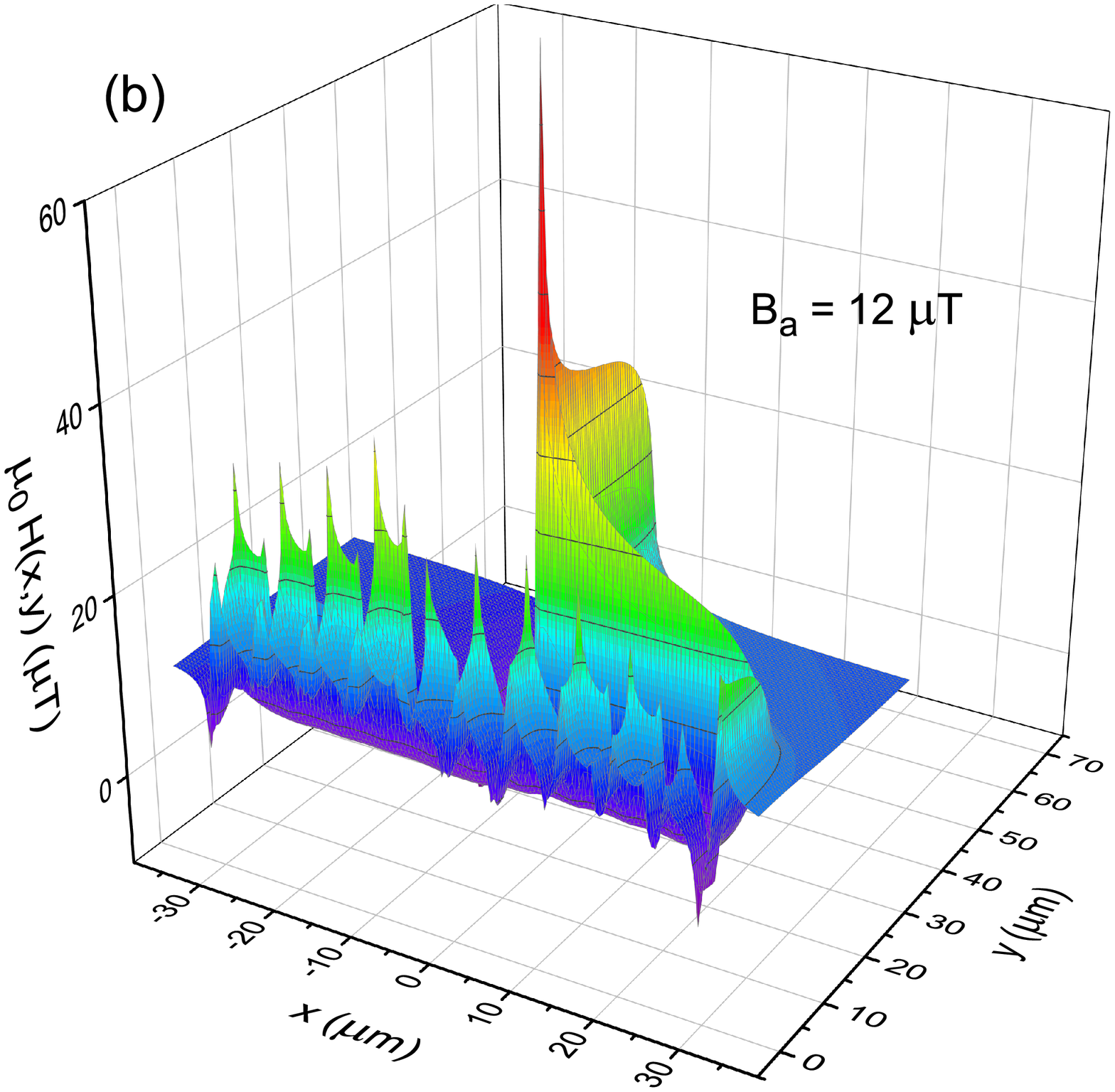}
\label{default}
\caption{Perpendicular total magnetic field $H(x,y)$ in the upper half ($y > 0$) of the $N$ = 11 parallel SQUID array at time $\tau = 9000$ for (a) $B_a = 0$ and (b) $B_a$ = 12 $\mu$T.}\end{center}
\end{figure}

Figures 6(a) and (b) show the calculated perpendicular total magnetic field $H(x,y)$, Eq. (9) and (14), inside and outside of the $N = 11$ parallel SQUID array ( $y \ge 0$) for $B_a = 0$ and $B_a$ = 12 $\mu$T at time $\tau$ = 9000. Strong magnetic field enhancements are visible along the edges of the bias-lead and the upper edge of the busbar, with a particularly strong field at the corners between bias-lead and busbar due to strong current crowding. The magnetic fields around and inside of the holes look complicated and change with time as the junction currents oscillate with time. At  $B_a$ = 12 $\mu$T the applied magnetic induction adds to the total field and the Meissner shielding currents induced in the busbars and bias-leads generate additional magnetic fields along edges. The lowest magnetic field, in the centre region of the busbar (not visible here), is about 3.5 $\mu$T$/ \mu_0$ and thus the maximum busbar-shielding is about 70$\%$.
\\
\par

\begin{figure}[h]\begin{center}
\includegraphics[width=0.57\textwidth]{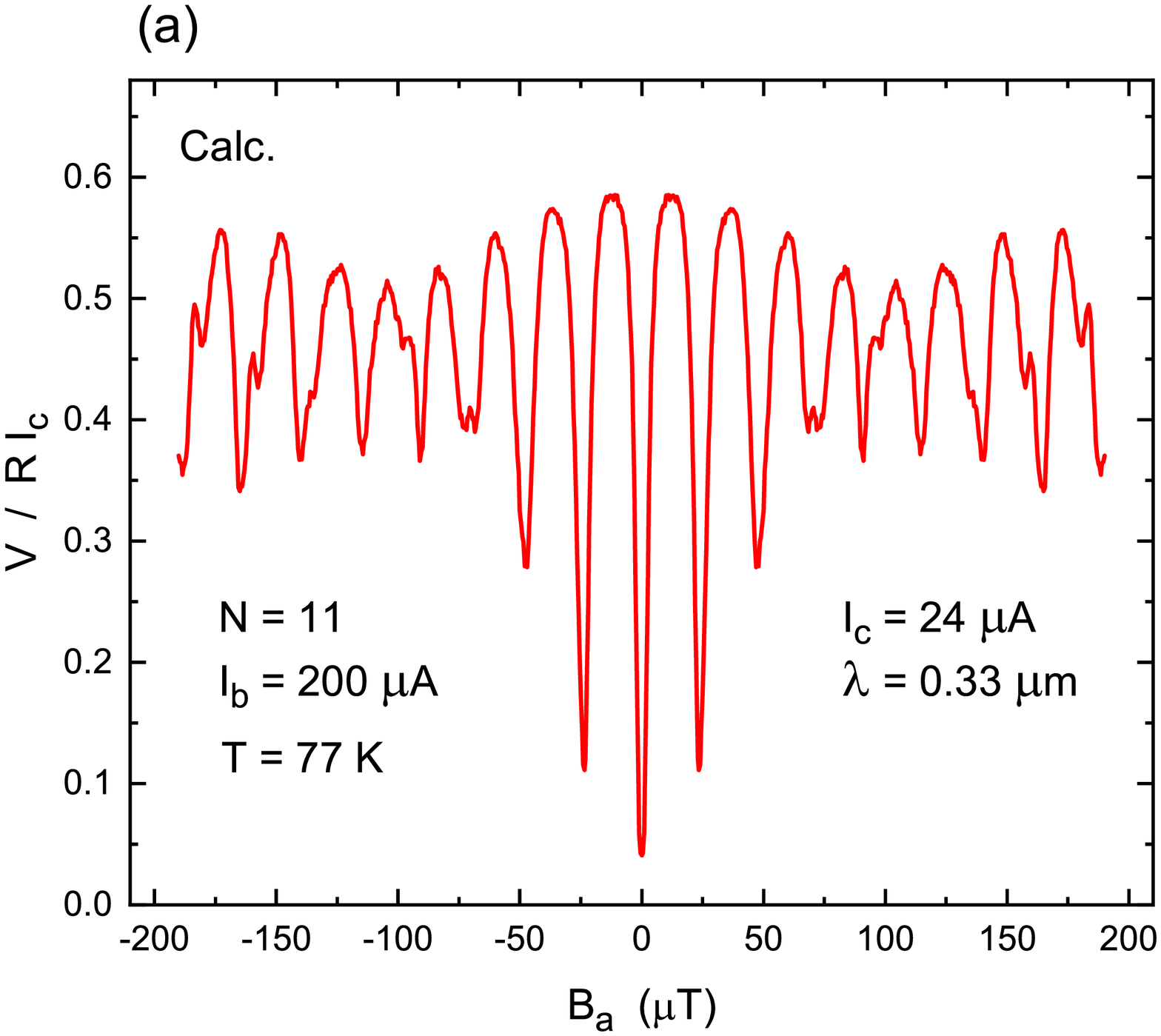}
\includegraphics[width=0.57\textwidth]{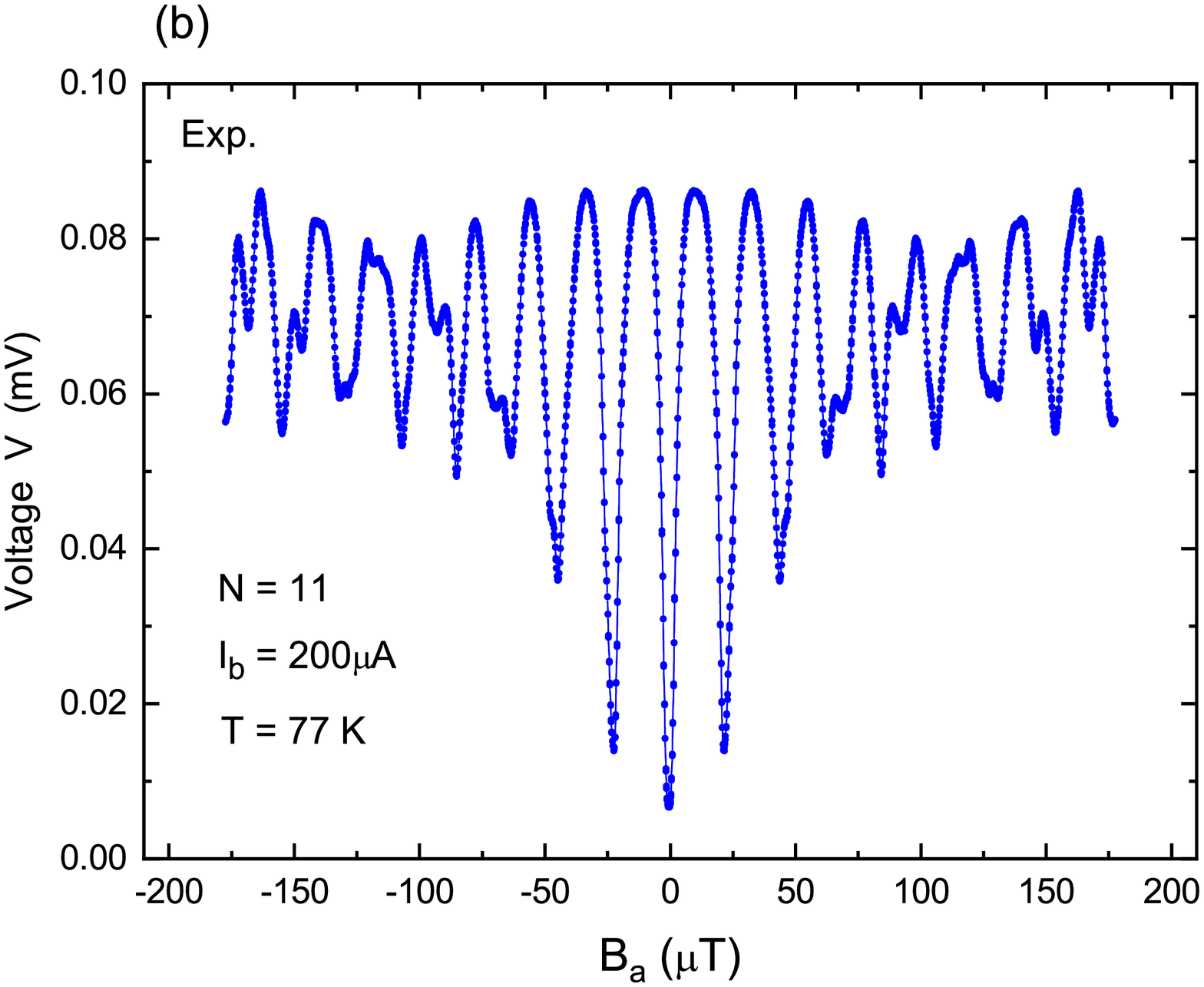}
\caption{Responses $V$ versus $B_a$ at $T$ = 77 K for (a) calculation and (b) experiment. }
\label{default}
\end{center}
\end{figure}

Figure 7(a), which is the most important result of our paper, shows the calculated time-averaged voltage $V$ versus the applied perpendicular magnetic induction $B_a$, from -200 $\mu$T to +200 $\mu$T, for the $N = 11$ parallel SQUID array at a bias current $I_b$ = 200 $\mu$A, assuming zero $I_{ck}$, $R_k$ spreads. The calculation includes Johnson noise at $T$ = 77 K. The average junction critical current density $I_c$ and the London penetration depth $\lambda$, needed as input parameters in our calculation, were chosen to give the best agreement with our experimental data displayed in Fig. 7(b). These parameters are $I_c$ = 24 $\mu$A and $\lambda$ = 0.33 $\mu$m. The temperature dependence of the London penetration depth $\lambda(T)$ for YBCO thin-films has been widely investigated \cite{TAL15}, and Chen {\it{et al.}} \cite{CHE14} found for high quality YBCO thin-films, using AC-susceptibility measurements, $\lambda(77K) \approx 0.3 \, \mu$m, which is similar to our value. As can be seen, our calculation in Fig. 7(a) agrees very well with our experimental data shown in Fig. 7(b). The calculation reproduces accurately the experimental ratio of maximal to minimal voltage as well as the overall experimental envelope modulation. Also, the dips appear at the correct $B_a$ values. In addition, the shoulder-peak that initially appears near the second side minima and then propagates outwards with increasing $B_a$ is closely reproduced. 
In the experiment the bias-current $I_b$ = 200 $\mu$A was chosen to maximise the transfer function $d V / dB_a$ around the centre dip. From the $I_c$ value it follows that $I_b = 0.758 \, N \,I_c$. In cases where $I_b < N \, I_c$, it is important to take the effects of Johnson noise in the calculation fully into account, as we have done here. Comparing the experimental maximum voltage with the calculated maximum $V/(R I_c)$ value in Fig. 7(a), one obtains an average single junction resistance $R$ = 6.2 $\Omega$. Some of the tiny spikes around the upper parts of the calculated curve are due to numerical inaccuracies of a finite temperature calculation due to a limitation in available computation time.
\\
\par

\begin{figure}[h]\begin{center}
\includegraphics[width=0.56\textwidth]{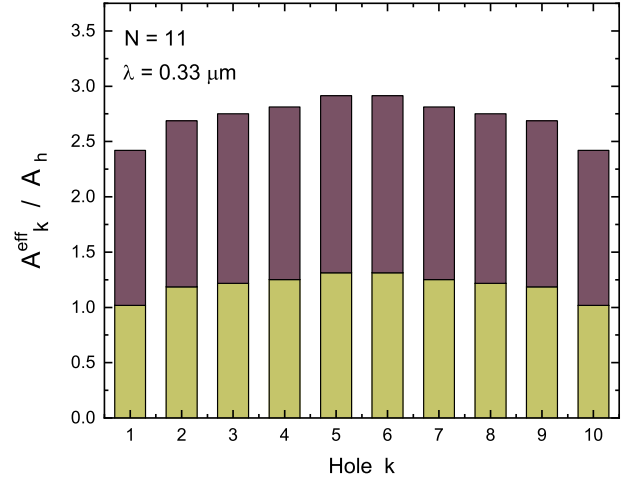}
\caption{Effective area enhancement factor $A^{eff}_k  /  A_h$ of Eq. (40) versus the hole index $k$ for $N=11$. The average enhancement factor is 2.72. The lower part of the bars is the $\mu_0 \lambda^2 \oint \bm{j} \cdot d\bm{l}$ fluxoid contribution to the effective area (Eq. (39)) while the upper part is the $\Phi_k$ fluxoid contribution. }
\label{default}
\end{center}
\end{figure}

Our calculation reveals that the appearance of a broad envelope modulation in the $V$ versus $B_a$ response shown in Figs. 7(a) and (b) is due to inhomogeneous fluxoid focussing where the effective areas $A_k^{eff}$ of the two holes ($k=5$ and $k=6$) closest to the centre are larger than the effective areas of the holes at the ends. The values of $A_k^{eff}$, with $k=1,..., 11$, calculated from Eq. (40), are displayed in Fig. 8. The lower part of the bars is the $\mu_0 \lambda^2 \oint \bm{j} \cdot d\bm{l}$ fluxoid contribution to the effective area (Eq. (39)) while the upper part is the $\Phi_k$ fluxoid contribution (which contains the applied flux). 
The average effective area enhancement is $A_{av}^{eff} / A_h$ = 2.72. Using this enhancement factor, one obtains for the first side minimum position $B_{a,0}  := \Phi_o / A_{av}^{eff}$ = 23.78 $\mu$T, in close agreement with both Figs. 7(a) and (b). The effective area difference of 17\% between the end and the centre holes is responsible for the strong envelope modulation seen in Figs. 7(a) and (b). The complicated interference pattern seen in Fig. 7(a) is very sensitive to the actual form of the effective area $A_k^{eff}$ distribution (Fig. 8). In a simple lumped-element simulation an envelope modulation can also be produced by varying the geometric area sizes of the holes \cite{MIT19}. The effect of varying the geometrical areas in parallel SQUID arrays on the $V$ versus $B_a$ response has been investigated in detail by Oppenl\"{a}nder {\it{et al.}} \cite{OPP01} who simulated the behaviour of SQIF's using a lumped-element approach. It is important to note that these lumped-element simulations cannot properly account for the effect of flux focussing and Meissner shielding-currents flowing in wide tracks, busbars and leads, and in particular are not suitable to calculate how the injected current from a wide bias-lead fans out into the junctions currents. Lumped-element simulations are thus unable to accurately describe the $V$ versus $B_a$ response of a parallel SQUID array with wide superconducting thin-film busbars and leads.
\\
\par

\begin{figure}[h]\begin{center}
\includegraphics[width=0.56\textwidth]{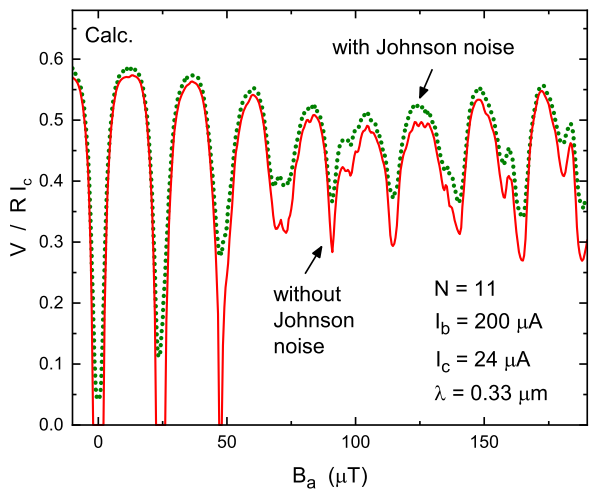}
\caption{Comparison of calculated response $V$ versus $B_a$ with Johnson noise (dashed green curve, Fig. 7(a)) and without Johnson noise (solid red curve).}
\label{default}
\end{center}
\end{figure}

Figure 9 demonstrates the importance of taking Johnson noise of the junction resistors into account, when calculating the $V$ versus $B_a$ response. The solid red curve in Fig. 9 was obtained without Johnson noise and shows sharper dips and narrow regions of zero voltage in contrast to the dashed green curve which is identical to Fig. 7(a) which includes Johnson noise. This kind of response difference is well known from early simulations for a SQUID with $N$ = 2 \cite{TES77}.
We found that numerical calculations for the $N$ = 11 parallel SQUID array that include Johnson noise are computationally about 30 times more demanding than calculations without Johnson noise. At $T = 77$ K, particularly at small average voltages $V$, one has to calculate the time-evolution of the junction phase differences $\varphi_k(\tau)$ over a very long time period $\tau$ in order to obtain a statistically accurate time-averaged voltage.  The accuracy of calculations shown in Fig. 7(a) is about 2{\%}.
\\
\par

\begin{figure}[h]
\begin{center}
\includegraphics[width=0.56\textwidth]{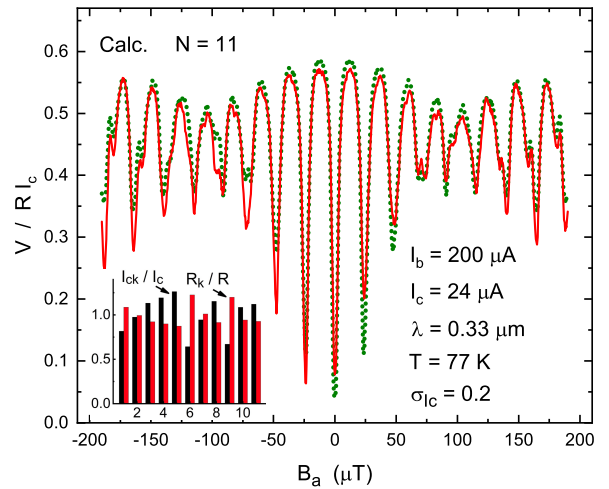}
\caption{Effect of $I_{ck}$, $R_k$ spreads with $\sigma_{Ic}$ = 0.2 (full red curve) on $V$ versus $B_a$. The green dotted curve is for no spread and is identical to Fig. 7(a). The inset shows the Gaussian random $I_{ck} / I_c$ and $R_k / R$ used.}
\label{default}
\end{center}
\end{figure}

Up to this point our calculations have not considered spreads in critical junction currents $I_{ck}$ and junction resistances $R_k$. The effect of $I_{ck}$, $R_k$ spreads with a standard deviation of
$\sigma_{Ic}$ = 0.2 is shown in Fig. 10. Here we assume that $I_k$ and $R_k$ are anti-correlated according to the empirical law $R_k I_{ck} \propto J_c^{1/2}$ \cite{GRO97} where $J_c$ is the junction critical current density which is assumed to be a constant. The Gaussian random $I_{ck} / I_c$ and $R_k / R$ values that were used are shown in the inset of Fig. 10. As can be seen, compared to the case of no spread (dotted green curve; identical to Fig. 7(a)), the spread (full red curve) breaks the symmetry about the $V$ axis and $V(B_a) \neq V(-B_a)$. The envelope modulation is only mildly affected and the previously mentioned shoulder peaks stay at their positions. In contrast, the experimental data in Fig. 7(b) shows symmetric behaviour which is due to our experimental procedure which averages the voltages $V$ for direct and reversed bias-current $I_b$, in order to cancel any apparatus voltage offset. Thus the measurement procedure symmetrised the $V$ vs $B_a$ curve in Fig. 7(b). The fact that the calculation in Fig. 7(a) agrees so well with the symmetrised experimental data in Fig. 7(b) indicates that the $I_{ck}$, $R_k$ spreads of our experimental $N = 11$ parallel SQUID array must be smaller than 0.2.
\\
\par

\begin{figure}[h]\begin{center}
\includegraphics[width=0.56\textwidth]{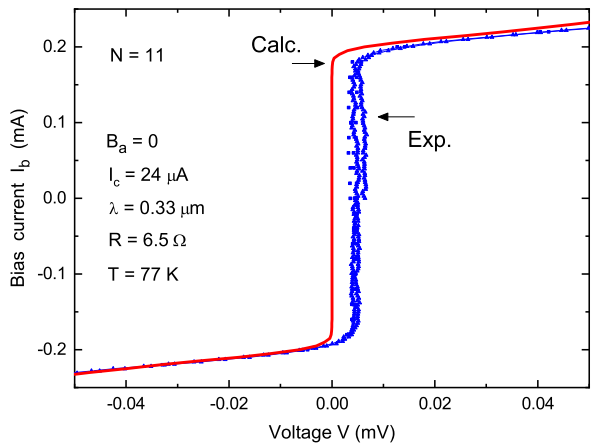}
\caption{Comparison between the experimental and calculated bias-current $I_b$ versus the time-averaged voltage $V$ across the $N$ = 11 parallel SQUID array for $B_a$ =0. In the calculation the resistance of each JJ is assumed to be $R$ = 6.5 $\Omega$.}
\label{default}
\end{center}
\end{figure} 

The time-averaged voltage $V$ at $B_a$ = 0 in Figs. 7(a) and (b) is dependent on the bias current $I_b$. Figure 11 displays the calculated and experimental $I_b$ versus $V(B_a=0)$ dependence. The calculation included Johnson noise at $T$ = 77 K and used the same parameters as in Fig. 7(a), {\it{i.e.}} $I_c$ = 24 $\mu$A and $\lambda$ = 0.33 $\mu$m, again assuming zero $I_{ck}$, $R_k$ spreads. Figure 11 shows that a junction resistance $R$ = 6.5 $\Omega$ fits quite well the experimental data and the thermal rounding seen in the experimental $I_b$ versus $V$ curve is well reproduced. The JJ resistance value $R$ = 6.5 $\Omega$ is close to $R$ = 6.2 $\Omega$ which was extracted above from Figs. 7(a) and (b). As seen in Fig. 11, the experimental curve has a voltage offset error of 5 $\mu$V found to be caused by a systematic error from our measuring apparatus. 
\\
\par
From the supercurrents that are flowing parallel and very close to the junctions we can calculate the fluxoids in the junctions. We find that for the largest magnetic induction $B_a$ = 190 $\mu$T, the value for the fluxoids in junctions is about 0.06 $\Phi_0$. Thus, the critical current of JJs is not affected by the applied magnetic field which is sufficiently small so that our initial assumption of a nearly constant critical current density across junction areas is well justified.
\\
\par

\begin{figure}[htbp]\begin{center}
\includegraphics[width=0.56\textwidth]{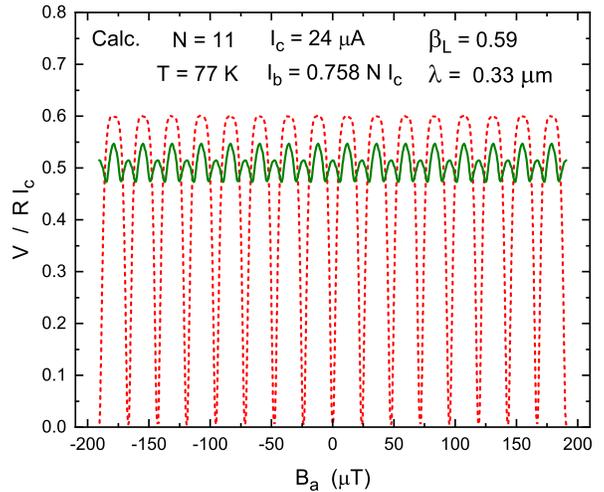}
\caption{$V$ versus $B_a$ using the lumped-element model with Johnson noise. The dashed red curve corresponds to homogeneous bias-current injection and the full green line represents central bias-current injection. The SQUID-loop screening parameter used in the calculation is $\beta_L $ = 0.59 .}
\label{default}
\end{center}
\end{figure}

It is interesting to compare the results from our comprehensive model with the simpler lumped-element model \cite{MIL91, OPP01}. In the lumped-element approach one also solves a system of coupled differential equations for the junction phase differences, but these equations contain partial inductances for all the tracks along which currents are flowing. To obtain the set of coupled equations for the phase differences one has to utilise Kirchhoff's law for all current vertices. Thus, in a lumped-element model, currents fanning out from the bias-current injection lead into wide busbars cannot be modelled properly. Therefore, in the lumped-element approach, one is often faced with the problem of how to best simulate the injection of the bias-current \cite{MIT19}. It is clear that the lumped-element model only works well if the device is made up of very narrow tracks. Figure 12 shows the calculated result using the lumped-element model \cite{MIT19} with Johnson noise for an $N = 11$ parallel SQUID array assuming very narrow tracks. In order to allow a fair comparison with our more elaborate model, we artificially included a constant fluxoid focussing factor of 2.72 (Fig. 8) and chose for the screening parameter $\beta_L = 2 I_c L_s / \Phi_0$ = 0.59 which was obtained by extracting the average self-inductance $L_s$ per array hole from our comprehensive model. The average $L_s$ was determined by forcing a loop current to flow around individual holes and then evaluating the corresponding hole fluxoid. This showed that the self-inductances of the outside loops are about 2.3\% larger than the inner ones. The dotted red curve in Fig. 12 shows the result for homogeneous bias-current injection \cite{MIT19} where all the top array current vertices receive the same bias-current $I_b / N$. In contrast, the solid green curve in Fig. 12 shows the calculated result for central injection, where the full bias-current $I_b$ is injected into the top centre current vertex, {\it{i.e.}} above the central ($k = 6$) junction. It can be seen that the lumped-element model very strongly depends on the bias-current injection scheme chosen and completely fails to reproduce our experimental data displayed in Fig. 7(b). Our comprehensive model, due to its greater details, takes a factor of about 10$^4$ more computation time than the computationally fast lumped-element model.
\\
\par
We have also fabricated and measured 8 other parallel SQUID arrays with numbers of JJs ranging from $N$ = 4 to $N$ = 81 \cite{MIT19} and we intend to compare these experimental data with our comprehensive model in an upcoming paper.

\section{Summary and Conclusion}
In this paper we have presented a comprehensive theoretical model that allows us to calculate with high accuracy the magnetic field response of an $N$ = 11 parallel SQUID array with wide thin-film geometric structures, operated in the voltage state. The model calculates the fluxoids for each SQUID array hole during the time-evolution of the phase differences of the JJs. This was achieved by solving numerically the second-order linear Fredholm integro-differential equation for the stream function, derived from the second London equation and Biot-Savart's law, with boundary conditions that are updated for each time step. The fact that the London penetration depth of YBCO thin-films at 77 K is greater than the thickness of our thin-film, allows us to solve the intero-differential equation in 2D. The Josephson equations and the second Ginzburg-Landau equation for the phase differences lead to a system of coupled first-order nonlinear differential equations which depend on the stream function which describes the time-varying supercurrent density within the thin-film array structure. The equations also take into account the Johnson noise from the JJs. Compared to the much simpler and far less accurate lumped-element model approach, our comprehensive model, while computationally much more demanding, leads to highly accurate predictions.\\
\par
We have tested the predictive power of our model by comparing our model results with our experimental data for an $N$ = 11 parallel SQUID array with wide thin-film structures. The theoretical model requires only two parameters, {\it{i.e.}} the junction critical current density $I_c$ and the London penetration depth $\lambda$. Both parameters were adjusted to give the best allover agreement with the experimental $V(B_a)$ curve of the array. The model predicts with unprecedentedly high accuracy the $V(B_a)$ curve of the array over a wide applied magnetic field range and also describes well the experimentally observed thermal rounding of the $I_b$ versus $V(B_a=0)$ curve. The model reveals that the observed envelope modulation of the experimental $V(B_a)$ curve is due to the non-equal effective hole areas of the array, where the centre holes have a larger effective area. Spreads in $I_{ck}$ and $R_{k}$ lead to $B_a$-asymmetry of the $V(B_a)$ curve. For wide thin-film geometric structures, the lumped-element model fails to correctly predict the $V(B_a)$ array response, because it cannot describe fluxoid focussing and in particular fails to appropriately handle the bias-current injection.

\begin{acknowledgments}
The authors would like to thank Colin Pegrum, J\"{o}rn Beyer, Marc Gali Labarias, Philip Fairman, Chris Lewis and John Clarke for stimulating discussions.
\end{acknowledgments}

\section{Appendixes}

\appendix

\section{Kernel $Q_F$}

The integral Eq. (5) is a convergent improper integral where the integrand is singular for $(x,y) = (x',y')$. In order to apply the method of integration by parts, one can smoothen the functions  $(y-y')/\sqrt{(c-x')^2+(y-y')^2}^3$ and  $(x-x')/\sqrt{(c-x')^2+(y-y')^2}^3$ in Eq. (5) by analytically integrating them over the small area of a square grid element $(\Delta x)^2$ size, so that these functions become continuously differentiable functions. This leads in Eq. (7) to a kernel of the form 

\begin{equation}
Q_F(x,y,x',y') = 
\end{equation}
\[
\frac{1}{(\Delta x)^2} \, \Bigg[ \, \Big[ \frac{\sqrt{\bar{x}^2 + \bar{y}^2}}{\bar{x} \, \bar{y}} \Big]_{x'-x -\Delta{x}/2}^{x'-x+\Delta{x}/2} (\bar{x}) \Bigg]_{y'-y -\Delta{x}/2}^{y'-y+\Delta{x}/2} (\bar{y}) \; \; 
\] \\
with the definition
\begin{equation}
\Bigg[ \, \Big[ f(x,y) \Big]_{s_1}^{s_2} (x) \Bigg]_{s_3}^{s_4} (y) \, = 
\end{equation}
\[
f(s_2,s_4) - f(s_1,s_4) - f(s_2,s_3)+f(s_1,s_3) \; .
\] \\
We also have tried a method suggested by Brandt \cite{BRA05} in order to avoid the singularity in Eq. (5). Brandt \cite{BRA05} uses the fact that $1 / \sqrt{(x-x')^2 + (y-y')^2}^{\,3}$ in Eq. (5) can be interpreted as $4 \pi$ times the magnetic field in the $x y$ plane of a point dipole of unit strength, positioned at $(x',y')$ and oriented in $z$ direction. Since the magnetic flux through the infinite $x y$ plane is zero, this leads to an additional equation that Brandt \cite{BRA05} uses to eliminate any unphysical singularity. We found that this method is less accurate than our method described above. 

\section{Analytical expressions for the functions $P^u_0(x,y)$ and $P^u_k(x,y)$  }

Using Eqs. (18) and (19), the function $P^u_0(x,y)$ is obtained by integrating the line integral in Eq. (16) along the contours $\partial \Omega^L$, $\partial \Omega^R$ and $\partial \Omega^T$ (see Fig. 2), which results in
\begin{equation}
P^u_0(x,y) = \frac{1}{8 \pi} \left[ \; \alpha_L(x,y) + \alpha_R(x,y) + \alpha_T(x,y) \; \right] \; ,
\end{equation}
where,
\begin{equation}
\alpha_L(x,y) = \frac{1}{x+c} \frac{\tilde{y}}{\sqrt{(x+c)^2 
+\tilde{y}^2}} \, \bigg\rvert_{\tilde{y}=b-y}^{b+l-y}\, 
\end{equation}
\[
- \, \frac{1}{y-b} \frac{\tilde{x}}{\sqrt{\tilde{x}^2 +(y-b)^2}} \, \bigg\rvert_{\tilde{x}=-a-x}^{-c-x}
\]
\[
+ \,  \frac{1}{x+a} \frac{\tilde{y}}{\sqrt{(x+a)^2 +\tilde{y}^2}} \, \bigg\rvert_{\tilde{y}= -y}^{b-y}\; ,
\]
with $c$, $b$, $l$ and $a$ defined in Fig.1,
and
\begin{equation}
\alpha_R(x,y) = \frac{1}{x-c} \frac{\tilde{y}}{\sqrt{(x-c)^2 +\tilde{y}^2}} \, \bigg\rvert_{\tilde{y}=b-y}^{b+l-y}\, 
\end{equation}
\[
+ \, \frac{1}{y-b} \frac{\tilde{x}}{\sqrt{\tilde{x}^2 +(y-b)^2}} \, \bigg\rvert_{\tilde{x}=c-x}^{a-x}
\]
\[
+ \,  \frac{1}{x-a} \frac{\tilde{y}}{\sqrt{(x-a)^2 +\tilde{y}^2}} \, \bigg\rvert_{\tilde{y}= -y}^{b-y}\; ,
\]
and
\begin{equation}
\alpha_T(x,y) = \frac{1}{c} \, \left[ \, \frac{x \,\tilde{x}}{y-(b+l)} - (y-(b+l)) \, \right] 
\end{equation}
\[
\frac{1}{\sqrt{\tilde{x}^2 + (y - (b+l))^2}} \, \bigg\rvert_{\tilde{x}=-c-x}^{c-x} \; .
\] \\
\par

The function $P^u_k(x,y)$, where $k=1, ..., N-1$, is obtained by integrating the line integral in Eq. (16) along the contour $\partial \Omega_k$ (Fig. 2) for $y' \geq 0$, which results in
\begin{equation}
4 \pi \, P_k^u(x,y) = -  \, \frac{\tilde{x}}{(y-h) \sqrt{\tilde{x}^2 + (y-h)^2}} \, \bigg\rvert_{\tilde{x}={x_k}-x}^{x_k+w_h-x}
\end{equation}
\[
 + \, \frac{\tilde{y}}{(x-x_k) \sqrt{(x-x_k)^2 + \tilde{y}^2}} \, \bigg\rvert_{\tilde{y}=-y}^{h-y}
\]
\[
-\,  \frac{\tilde{y}}{(x-(x_k+w_h)) \sqrt{(x-(x_k+w_h))^2 + \tilde{y}^2}} \, \bigg\rvert_{\tilde{y}=-y}^{h-y} \; ,
\] \\
where $x_k : = k \, (w_J + w_h) - w_h - a$ \, .

\section{Laplacian}

The superconducting thin-film 2D domain $\Omega^u$ (Fig. 2) is divided into small square aerial elements $w = (\Delta x)^2$ where the square grid spacing $\Delta x$ was chosen as $0.5 \, \mu m$ or $1.0 \, \mu m$. The grid points $\bm{r}_n$, with $n=1,...,N_g$, lie in the centre of these grid elements. \\
\par
In Eq. (24), if a point $\bm{r}_n$ lies more than a distance $\Delta x / 2$ from the boundary $\partial \Omega^u$ (which includes the boundaries $\partial \Omega_k$) or from the  junction boundary, then the Laplacian on the square grid operates such that $\sum_m \Delta_{nm} \, g_m = [ \, g(x_n + \Delta x, y_n) + g(x_n, y_n + \Delta x) - 4 \, g(x_n,y_n) + g(x_n - \Delta x, y_n) + g(x_n, y_n - \Delta x) \,]  / w$ where the sum over $m$ includes the four nearest neighbour sites of $\bm{r}_n$. \\
\par

In Eq. (24), if a point $\bm{r}_n$ lies only a distance $\Delta x / 2$ from the boundary $\partial \Omega^u$ or a junction, one has to use the Laplacian for a non-equidistant grid. For example if $\bm{r}_n$ is on the right side of a boundary line that runs along the $y$ direction, then
\begin{widetext}
\begin{equation}
\partial^2{g} / \partial{x}^2  \, {\Big{\vert}}_{(x_n,y_n)} \approx  \;
g(x_n-h_1,y_n) \frac{2/h_1}{h_1+h_2} - g(x_n,y_n) \frac{2}{h_1 h_2} + g(x_n+h_2, y_n) \frac{2 / h_2}{h_1+h_2}\, ,
\end{equation}
\end{widetext}
\noindent \\
where in this case $h_1= \Delta x /2$, $h_2 = \Delta x$, and $g(x_n-h_1,y_n)$ is the stream function value on the boundary line. Equivalent equations for the Laplacian are used for other $\partial \Omega^u$ boundary lines and along junctions. Special attention has to be given to points $\bm{r}_n$ near corners.

\section{Johnson current noise}
We assume that the uncorrelated Johnson noise from the normal resistances of the junctions are the dominant noise sources, compared to junction shot noise or thermal fluctuations in critical currents \cite{TES77}. When solving Eqs. (34) and (35) numerically, the normalised noise currents $I_k^{Noise}/I_c$, with $k=1,..., N$, become sequences of random numbers corresponding to successive averages over small time steps $\Delta \tau$ of the continuous noise currents $I_k^{Noise}(\tau)/I_c$. Thus, each noise current is an independent Gaussian random variable with mean square deviation $2 \Gamma_k / \Delta \tau$ and zero average \cite{TES77,VOS81}, where $\Gamma_k$ is the effective noise strength
\begin{equation}
\Gamma_k = \frac{2 \pi k_B T}{I_{ck} \Phi_0} \;,
\end{equation}
and $k_B$ is the Boltzmann constant and $T$ the temperature the device is held at (here $T$= 77 K).
When applying the above concept to a single resistively shunted JJ, our numerical results agree with the Fokker-Planck calculations of Ambegaokar and Halperin \cite{AMB69}.\\
\par
\bibliography{busbar}

\end{document}